\documentclass[12pt]{article}
\usepackage{amsfonts}
 \usepackage{hyperref,setspace,bbm}
\usepackage[font=small, margin=1.5cm]{caption}

\hypersetup{%
  colorlinks=false,
  pdfborder = {0 0 0.5 [3 0]}
}
\usepackage{breakurl}

\usepackage{multirow}
\usepackage[left=0.8in,right=0.8in,top=1in,bottom=1in]{geometry}
\usepackage{amsmath}
\usepackage{xcolor}
\usepackage{natbib}
\usepackage{graphicx}
\usepackage{subfigure}
\usepackage{epstopdf}
\usepackage{ae,aecompl}
\usepackage{mathtools} 
\mathtoolsset{showonlyrefs=true}  
\usepackage{bm}
\usepackage[multiple]{footmisc} 
\usepackage{amsthm}

\newtheorem{proposition}{Proposition}
\newtheorem{corr}{Corollary}

\usepackage{mathtools} 
\mathtoolsset{showonlyrefs=true}
\DeclareMathOperator*{\argmin}{arg\,min}
\DeclareMathOperator*{\essup}{ess\,sup}

\def\Q{{\mathbb Q}} %

\def\1{1{\hskip -3.3 pt}\hbox{I}}

\numberwithin{equation}{section}
\numberwithin{theorem}{section}

\title{Understanding the Non-Convergence of Agricultural Futures via Stochastic Storage Costs and Timing Options}
\date{\today}
\author{Kevin Guo\thanks{\small{Industrial Engineering \& Operations Research Department, Columbia University, New York, NY 10027. E-mail: \mbox{klg2138@\,columbia.edu}.}} \and Tim Leung\thanks{\small{Department of Applied Mathematics, University of Washington, Seattle WA 98195. E-mail:
\mbox{timleung@uw.edu}. Corresponding author. } }}

\begin{document}

\maketitle
\abstract{This paper studies the market phenomenon of non-convergence between futures and spot prices in the grains market.   We postulate that the positive basis observed at maturity stems from the futures holder's timing options to exercise the shipping certificate delivery item and subsequently  liquidate the physical grain.  In our proposed approach, we  incorporate  stochastic spot price and storage cost, and solve an optimal double stopping problem to give the optimal strategies to exercise and liquidate the grain. Our new models for stochastic storage rates lead to explicit no-arbitrage prices for the shipping certificate and associated futures contract. We calibrate our models to empirical futures data during the periods of observed non-convergence, and illustrate the  premium generated by the shipping certificate.}

\vspace{10pt}

\newpage
\section{Introduction}\label{Introduction}

\begin{figure}[t]
    \centering
    \includegraphics[width=3.1in,trim=0.7cm 1cm 1.1cm 0.4cm, clip=true]{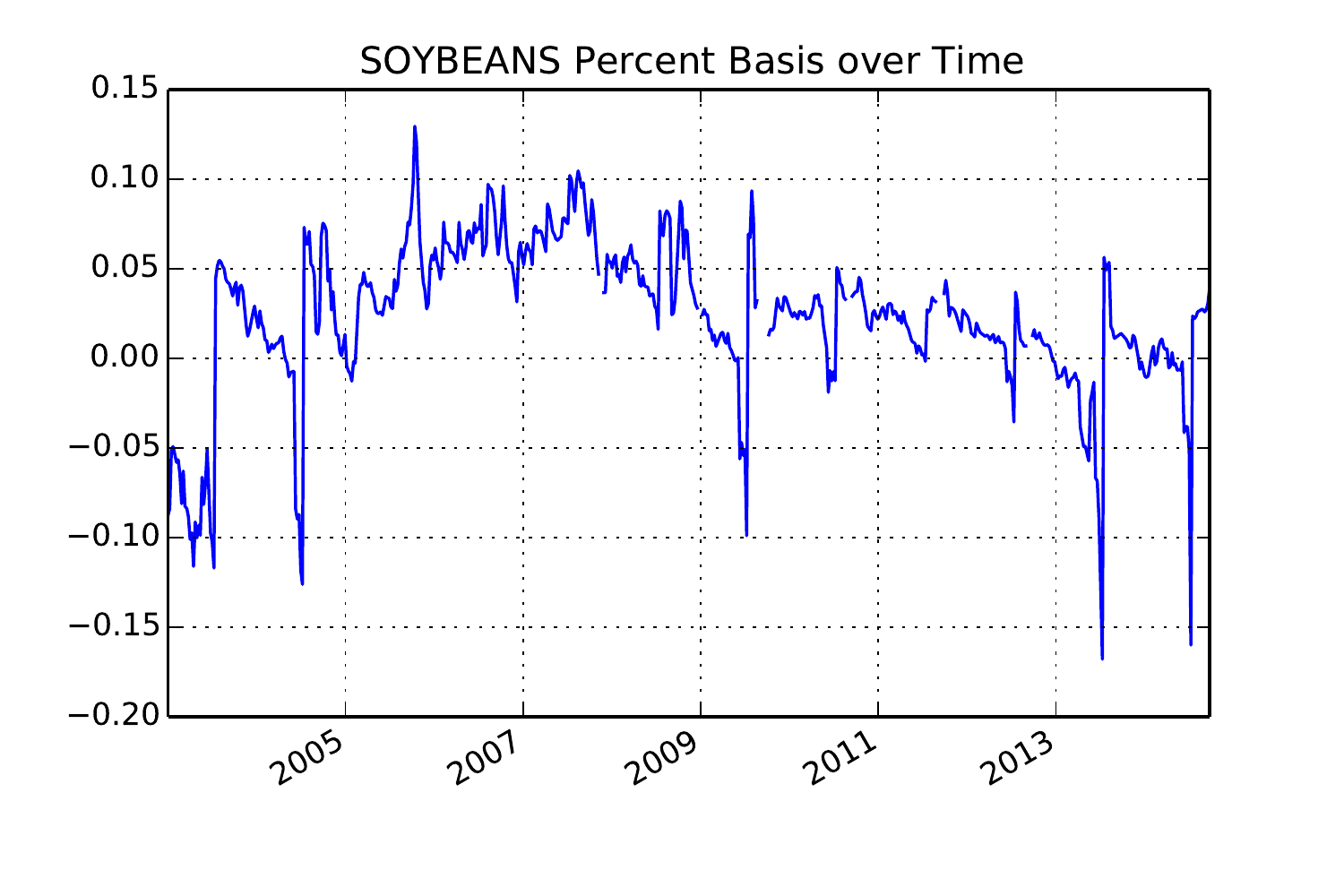}
    \includegraphics[width=3.1in,trim=0.7cm 1cm 1.1cm 0.4cm, clip=true]{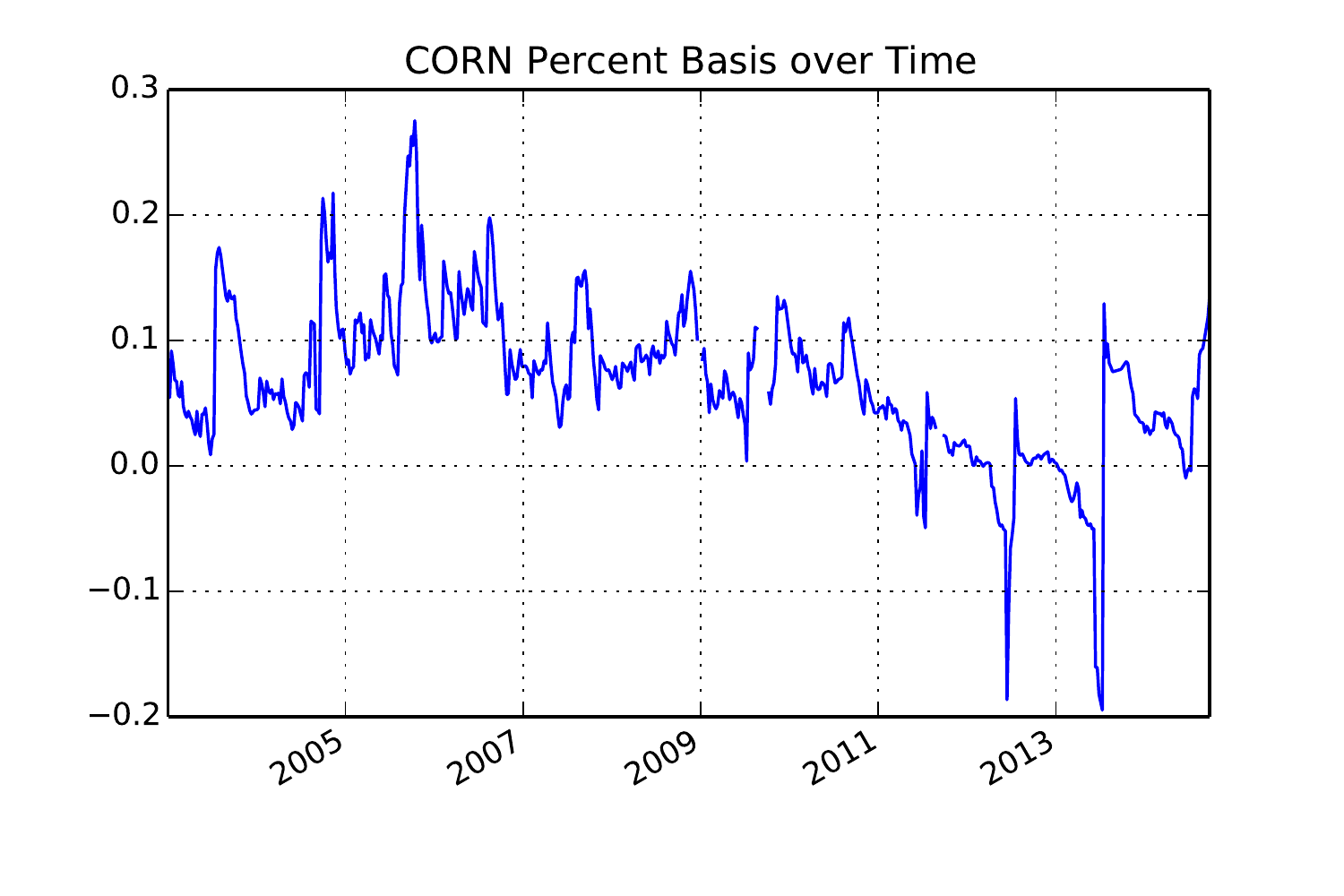}
    \caption{\small{Time series of basis for soybeans (left) and corn (right) futures. During 2004-2009,  the expiring futures price tends to be significantly higher than the spot price.}}
    \label{fig:compare}
\end{figure}

Standard no-arbitrage pricing theory asserts that spot and futures prices must converge at expiration. Nevertheless, during  2004-2009 traders observed significantly higher expiring futures prices for corn, wheat, and soybeans on the CBOT compared to the spot price of the physical grains. As shown in Figure \ref{fig:compare},  the unprecedented differential between cash and futures prices   reached its apex in 2006, where at the height of the phenomenon, CBOT corn futures had surpassed spot corn prices by almost 30\%! In the literature,  \cite{CTFC13} and \cite{Fishe11} reported  that  on July 1, 2008, the price for a July 2008 CBOT wheat futures contract closed at \$8.50 per bushel. On the other hand, the corresponding cash price in the Toledo, Ohio delivery market was only \$7.18 per bushel, a price differential of \$1.32/bu (+15\%). \cite{Irwin09} first coined the term ``non-convergence" for this phenomenon of observed positive premium, which recurred persistently from 2004 onwards. According to their study, ``performance has been consistently weakest in wheat, with futures prices at times exceeding delivery location cash prices by \$1.00/bu, a level of disconnect between cash and futures not previously experienced in grain markets."


However, a small difference between expiring futures and cash prices does not necessarily imply a market failure. Before expiration, futures and cash prices may differ due to the convenience yield, storage costs, or financing costs. Upon expiration, if cash prices are lower/higher than futures prices, then   arbitrageurs may profit from trading simultaneously in the spot and futures markets.  If sufficient numbers of arbitrageurs   engage in these trades, they will drive cash and futures prices to convergence at expiration. In fact, the futures expiration date and delivery date may also differ. After the last trade date, the exchange contacts the longest outstanding long who is notified of his obligation to undertake delivery. Before the month's end, the delivery instrument is then exchanged at the settlement price between long and short parties. Therefore, since  the delivery process does not occur immediately after the last trade date, cash and futures prices might still differ by a spread called the \emph{basis}. In this paper, we use the following definition for the basis:  \[\text{basis = futures price - spot price}.\] 
Figure \ref{fig:compare} displays the basis   time series  associated with the expiring futures on  soybeans and corn.

 Since the short party  may choose the location and time to deliver, \cite{Biagini07} posit that futures price should be biased below the spot price on the last trade date.  However, their theoretical model would yield  the opposite of the empirical observations in the grain markets. In fact, the positive basis in the CBOT grain markets between 2004-2009 were too large to have been caused by the small inefficiencies of the delivery process. This motivates us to investigate the factors that drive the non-convergence phenomenon.  
 
In order to explain the positive premium, one  must turn to embedded $long$-$side$ options in the futures. Long-side options in futures markets depend totally on the idiosyncrasies of each commodity's exchange traded structure. The survey paper by  \cite{carmona13}  demonstrates the appropriate model for a commodity varies highly depending on storability, instantaneous utility, and alternatives. At expiration, a  CBOT agricultural futures contract  does not deliver the physical grains  but an artificial instrument called the \emph{shipping certificate} that  entitles its holder to demand loading of the  grains from a warehouse at any time. Before exercising the option to load, the holder must pay a fixed storage fee to the storage company,\footnote{Only a  small number of  storage companies that  have contracts with the futures exchange are allowed to issue shipping certificates. They  are also called the regular firms in the industry.} as stipulated in the certificate.    Since the storage capacity of grain elevators is limited and expensive, the number of grain elevators is fixed to a minimum necessary to efficiently carry out transfers of grain from one transport system to another.\footnote{See \url{http://www.cmegroup.com/rulebook/CBOT/II/11/11.pdf}} Thus, like a fractional-reserve banking system, shipping certificates alleviate the congestion of grain elevators by only keeping enough grain on hand to satisfy instant withdrawal demand. A detailed explanation on the structure of the shipping certificate market can be found in  \cite{Fishe11} and \cite{Irwin11}. 

In this paper, our  storage differential hypothesis posits that when the certificate storage rate is sufficiently low, investors will pay a premium for the certificate over the spot grain in order to save on storage cost over time, resulting in non-convergence of futures and spot prices. When the storage cost of the certificate is set lower than the true storage cost paid by the regular firm, the regular firm will cease to issue the unprofitable shipping certificates. Since   shipping certificates can only be issued by a set number of regular firms with  limited     inventories, the market cannot issue certificates with lower fixed storage rates to keep the market flowing. Instead, since the supply of certificates remains fixed, the value of existing shipping certificates will be bid up in the secondary market, resulting in a premium over the spot price. On the other hand, during periods where the certificate storage rate is set much higher than the market storage rate, the certificate should not command any premium over the spot because  agents would exercise and store at the lower market rate. Therefore, as shown by \cite{Fishe11}, large quantities of certificates remaining unredeemed under the storage differential hypothesis becomes a strong predictor of non-convergence. In fact, in 2009 under mounting evidence that storage differentials were responsible for non-convergence, the CBOT raised the certificate storage rate for wheat, after which non-convergence decreased significantly.\footnote{\url{http://www.cftc.gov/idc/groups/public/@aboutcftc/documents/file/reportofthesubcommitteeonconve.pdf}} This observation is consistent with our findings in this paper. 

Let us point out an alternative explanation for non-convergence even though it is not the approach in this paper.  The speculator hypothesis for non-convergence postulates that large $long$ positions held by commodity index traders (CITs) have made it impossible for arbitrageurs to carry sufficient grain forward to drive terminal prices to convergence (see \cite{Tang12} an example of the effects of excess speculation).  While the speculator hypothesis is plausible, empirical studies by \cite{Irwin11} and \cite{2Irwin11}    found no evidence that rolling  or initiation of large positions by index funds had contributed to an expansion of the basis. In this paper, we illustrate mathematically how the  non-convergence phenomenon can arise under rational  no-arbitrage models with stochastic storage rates.


We propose two new   models that incorporate the stochasticity of the market storage rate  and capture the storage option of the shipping certificate by solving two optimal timing problems, namely, to exercise the shipping certificate and subsequently liquidate the physical grain.   First, we propose the Martingale Model whereby the spot price minus storage cost is a martingale while the stochastic  storage rate follows an Ornstein-Uhlenbeck (OU) process. In addition, we present a second model in which the stochastic storage rate is a deterministic function of the spot price that follows an  exponential OU (XOU) process.  Among our results,  we provide  explicit  prices for the shipping certificate, futures prices, and the basis size under a two continuous-time no-arbitrage pricing models with stochastic storage rates.  By examining the divergence between expiring futures prices and corresponding spot prices, we derive the timing option generated by the differential between the   market storage rate   and the constant storage rate stipulated in the shipping certificate, which    explains the non-convergence phenomenon in agricultural commodity markets. We also fit our model prices to market data and extract the numerical value of the embedded timing option.

The rest of the paper is organized as follows. Section \ref{Literature Review} reviews the   literature on the subject of non-convergence for agricultural futures.  In Section \ref{GBM with Stochastic Storage}, we discuss a martingale spot  price model with an OU stochastic storage rate, and derive the  certificate price as well as the optimal exercise and liquidation timing strategies. In Section \ref{XOU with Local Storage}, we analyze a shipping certificate valuation  model with a local  stochastic storage rate and an exponential OU spot   price. Section \ref{Conclusion} concludes. Proofs are provided in the Appendix.

\section{Related Studies}\label{Literature Review}
 The long history of the  theory of storage dates back  to \cite{Kaldor1939} who argued that the future price should reflect the spot price plus storage cost via a no-arbitrage relationship. \cite{Johnson60} proposed an extension of Kaldor's model which related inventories and hedging motivations to the intersection of futures and spot markets. However, in order to account for possibly backwardated futures curves,  \cite{Brennan1958} and \cite{Working1948} developed the notion of a stochastic convenience yield, a fictitious dividend that  accrues to the commodity holder, but not the futures holder. Furthermore,   \cite{FamaFrench87} and \cite{Rowenhorst07} found plenty of empirical evidence for the theory of storage by examining inventories data. These authors not only created a theoretical basis for understanding commodity spot and futures prices, but also empirically demonstrated the validity of the theory of storage over a century. 
 
 

Much of the literature on embedded options in futures contracts studies  the short-side options which $lower$ the futures price below the spot price.  For example, \cite{jarrow05} estimate the  values of  the delivery option, which allows the short to choose the location of cheapest delivery. In addition, \cite{Biagini07} compute model-free futures prices for the short-side timing option. In contrast,  our models explain how the futures price can be $higher$ than the spot price at maturity.  Our proposed approach contributes to the theory of storage as it  provides a new link between the futures and spot markets through the storage cost differential and the associated timing option.  In a related study,  \cite{Hinz2010} consider the impact of storage cost constraints on commodity options, but their model cannot account for backwardated futures curves or  non-convergence.

\cite{Fishe11}  consider  an alternative model in which  non-convergence reflects the value of an exchange option due to the scarcity of shipping certificate. They incorporate  a  long-side option but not stochastic or differential storage rates.  The exchange option explanation is unsatisfactory because the exchange option is universal to all commodity futures, while the non-convergence phenomenon is observed  only in  the grain markets.   Our approach identifies different storage rates between the certificate and real world as the driver  for non-convergence at maturity. Finally, instead of using a closed-form approximation for the certificate price, we derive the explicit value of the shipping certificate   under different  stochastic storage rate models.

The storage differential hypothesis is supported by several   recent studies.   \cite{Irwin11} and \cite{CTFC13}  set up a discrete-time model and give  conditions for the number of shipping certificates in the market at equilibrium. While they  identified the difference between the market and certificate  storage rates as the crucial  factor for  non-convergence, they did not   compute the value of the shipping certificate. In this paper, we derive and compute explicitly the  prices of  the futures and  shipping certificate, and provide the necessary and sufficient  conditions for non-convergence. Furthermore, our  approach   requires  only the existence of the shipping certificate and no-arbitrage condition, and does not have specific assumptions on the characteristics of market agents and their interaction.

The core mathematical problem within our stochastic storage models is an optimal double stopping problem driven by a mean-reverting process. 
To this end, we adapt to our problem the results developed by  \cite{Leung14}  that study the optimal entry and exit timing strategies when the underlying is an   OU  process.  Other mean-reverting processes can also be used to model the market storage rate so long as the corresponding optimal double stopping problem can be solved analytically; see, e.g. \cite{LeungLi2015OU} and 
 \cite{LeungLiWang2014CIR} for the cases with  an  exponential OU and Cox-Ingersoll-Ross (CIR)  underlying, respectively, and related applications to futures trading in \cite{meanreversionbook2016} and \cite{Li2016}.

\section{Martingale Model with Stochastic Storage}\label{GBM with Stochastic Storage}
 We now  discuss a futures pricing model for a single grain type, with the spot price process $(S_t)_{t \geq 0}$. The cost of physical storage is stochastic,  represented by the rate process $(\delta_t)_{t \geq 0}$.  The spot price satisfies 
\begin{equation}
dS_t = (rS_t + \delta_t) dt + \sigma S_t dW_t,
\label{eq:S_t diffusion}
\end{equation}where  $r$ is the positive  risk-free rate,  $\sigma$ is the volatility parameter of spot grain, and $W$ is a standard Brownian motion under the risk-neutral measure $\mathbb{Q}$. In this  model, we assume that the commodity is continuously traded, with units of the commodity constantly being sold to pay the flat storage rate.  Hence,  the discounted spot  price $S_t$ \emph{net} storage cost,  i.e. $M_t :=  e^{-rt}S_t- \int_0^t \delta e^{-ru} du$, $t \geq 0$, is a $\mathbb{Q}$-martingale.



   Note that  $\delta_t$ is  the $net$ storage cost, which is the true storage cost minus the convenience yield associated with owning the physical grain. Furthermore, the storage rate $\delta_t$ is quoted in \$/bushel, and not as a proportion of the commodity price. In the standard treatment of storage rates, agents pay a proportion of the commodity price $S_t$ per bushel i.e. $\delta_t = c S_t.$ However, since empirical storage rates are   quoted in \$/bushel and not as a percentage of the crop, our specification of a flat storage rate $\delta_t$ is realistic and amenable for empirical analysis.  One can  view  the storage rate $\delta_t$ as a negative dividend rate  on the commodity which the commodity holder pays but the futures   holder does not. In our model,  the storage cost $\delta_t$ follows an Ornstein-Uhlenbeck (OU)  process
\begin{equation}
    d\delta_t = \kappa(\nu-\delta_t)dt + \zeta d\widetilde{W}_t,
    \label{eq:delta_t diffusion}
\end{equation}
 where $\widetilde{W}$ is a standard Brownian motion under $\Q$, and is independent of $W$.  The parameter $\kappa$ dictates the speed of mean-reversion for $\delta_t,$ $\nu$ is the average value of $\delta_t,$ and $\zeta$ is the volatility of the storage rate $\delta_t.$ The parameters are required to satisfy  the  condition  $2\kappa \geq \zeta.$ The filtration $\mathbb{F} \equiv(\mathcal{F}_t)_{t \geq 0}$ is generated by $(S_t)_{t\geq 0}$ and $(\delta_t)_{t\geq 0}$. We let $\mathcal{T}$ be the set of all $\mathbb{F}$-stopping times, and $\mathcal{T}_{s,u}$ be the  set of $\mathbb{F}$-stopping times bounded by $[s,u]$. 

At time $T$, the $T$-futures contract expires, and the long party receives the shipping certificate. This certificate is perpetually lived and gives  the holder an option  to load out the grain anytime, but  the holder pays  a constant storage rate $\widehat{\delta}$ before exercising this option. Note  that $\widehat{\delta}$ is a flat rate quoted in the futures contract, and thus must be positive. Since the certificate holder does not possess the physical grain,  and thus,  cannot derive any convenience yield from it. After exercising, the certificate holder then stores at the market rate $\delta_t$ until he chooses to liquidate the grain. We allow $\delta_t$ to be possibly negative to account for the convenience yield. The fixed costs, $c_1$ and $c_2$    respectively, are incurred upon  exercising and liquidation of the grain. 

The value of the shipping certificate can be obtained by solving two optimal timing problems. First, suppose  the agent has exercised at time $t \ge \tau$. The agent selects the optimal time  to liquidate the grains by solving\footnote{Throughout this paper, the shorthand notation ``$\sup$" stands for ``$\essup$". All computations in this paper are assumed to be under $\mathbb{Q},$ the risk-neutral measure.}
\begin{equation}
    J(S_t, \delta_t) = \sup_{\eta \in \mathcal{T}_{t,\infty}} \mathbb{E}\left[e^{-r(\eta-t)}\left(S_\eta-c_2\right) - \int_t^\eta \delta_u e^{-r(u-t)} du | \mathcal{F}_t \right].
	\label{eq:optimal liquidation GBM stoch}
\end{equation}
Working backward in time, the value function $J$ now serves  an input for the optimal exercise problem. The agent  receives the shipping certificate at time $T$, and selects the optimal time $\tau \geq T$ to exercise the grains. Therefore, the agent's value function at time $T$ is 

\begin{equation}
    V(S_T, \delta_T) = \sup_{\tau \in \mathcal{T}_{T, \infty}} \mathbb{E}\left[e^{-r(\tau-T)}\left(J(S_\tau, \delta_\tau)-c_1\right) - \int_T^\tau \widehat{\delta} e^{-r(u-T)} du | \mathcal{F}_T \right].
	\label{eq:optimal unloading GBM stoch}
\end{equation}

Economically, we interpret $J$ as the liquidation value of the commodity for an individual who optimally times storage, and $V$ as the price of the certificate (for an individual who can choose between storage rates). Furthermore, since $T=\tau=\eta$ is always a valid stopping time for \eqref{eq:optimal unloading GBM stoch}, we must have \[V(S_T, \delta_T) \geq S_T-c_1-c_2.\] In the absence of transaction costs, the shipping certificate is valued higher than the grain itself, and thus the certificate can be considered a long-side option. The value of the long-side option is quantified with the basis: if $T$ is the maturity of a futures contract, then the basis $w(S_T, \delta_T)$ is the difference between futures and cash prices at maturity 
\begin{equation}
w(S_T,\delta_T) = V(S_T,\delta_T) -S_T.\label{basis_def}
\end{equation}

As the shipping certificate, not the spot grain, is the true delivery instrument,   the price $F(t,S_t, \delta_t ;T)$ of a futures  contract expiring at $T$  satisfies the model-free price
\begin{equation}
    F(t,S_t, \delta_t ;T) = \mathbb{E}[V(S_T,\delta_T) | \mathcal{F}_t], \qquad t\le T.
		\label{eq:futures}
\end{equation} 
From this representation, it follows that the expiring futures price equals the certificate price $F(T,S_T, \delta_T; T) = V(S_T, \delta_T).$

Intuitively, the agent decides to liquidate when the spot price is sufficiently high. On the other hand, the agent may decide to exercise for two reasons: first if the spot price is sufficiently high, and second if the storage rate $\delta_t$ is sufficiently low relative to the certificate rate $\widehat{\delta}$. In the first case, the agent exercises and liquidates (i.e. $\tau^*=\eta^*$), and in the second case, he exercises the shipping certificate but  holds the commodity for longer, thus taking advantage of the lower storage rate $\delta_t$ until the eventual liquidation.  

The   stochastic storage rate $\delta_t$ is a crucial factor  for non-convergence. Indeed, if we instead consider  the simple   constant storage rate $(\delta_t \equiv \delta)$, then  the certificate pricing problem \eqref{eq:optimal unloading GBM stoch} simplifies to 
\begin{align}
\begin{split}
    V(S_T, \delta_T) = \sup_{\tau \in \mathcal{T}_{T,\infty}\, \eta \in \mathcal{T}_{\tau,\infty}} \mathbb{E}&\left[ e^{-r(\eta-T)}S_\eta - \int_{T}^{\tau} (\widehat{\delta} - (c_1+c_2)r) e^{-r(u-T)} du \right. \\
    &\left.- \int_{\tau}^{\eta} (\delta-c_2r) e^{-r(u-T)} du | \mathcal{F}_T \right]-c_1-c_2. 
\end{split}
\label{eq:second form}
\end{align}

 By inspecting the value function in  \eqref{eq:second form}, we see that at every time $t,$ the agent effectively has a choice between paying the storage rate $\widehat{\delta} - (c_1+c_2)r$ and the storage rate $\delta-c_2r.$ In other words, the agent will immediately lock in the lesser of the two rates. If $\delta-\widehat{\delta} \leq -c_1 r,$ then  the agent exercises immediately at expiration ($\tau^*=T$).   On the other hand, if $\delta-\widehat{\delta} > -c_1 r,$ then  the agent liquidates immediately after uploading ($\eta^*=\tau^*$). We recognize instantly that under the assumption of constant storage rates at least one stopping time ($\tau^*$ or $\eta^*$) is trivial. Either the agent exercises immediately, or he liquidates after exercising. In particular, this fact does not depend at all on the realized path  of $S$. Therefore, a constant storage rate model is insufficient since   the shipping certificate is never used for  storing   the grain   for a non-trivial period of time. Since certificate holders empirically store and exercise in a multitude of competitive markets, we must   consider a stochastic storage rate  $\delta_t$ in all our models. 

In order to solve for $J$ and $V$  under the dynamics \eqref{eq:S_t diffusion} and \eqref{eq:delta_t diffusion},  we need to study an ODE. Define the differential operator $\mathcal{L} \equiv \mathcal{L}^{a,b,c}$ by 
\begin{equation}
		\mathcal{L} = a(b-x)\frac{d}{dx} + \frac{1}{2}c^2 \frac{d^2}{dx^2},
		\label{eq:L}
\end{equation}
with the generic parameters $(a,b,c)$ with $a,c > 0$ and $b \in \mathbb{R}.$ This is the infinitesimal generator associated with an OU process. In turn, the ODE

\begin{equation}
		\mathcal{L}f(x)-rf(x) = 0
\end{equation}

\noindent has the two general classical solutions (see e.g. \cite{Borodin02})
    \begin{equation}
        H(x; a,b,c) = \int_0^\infty v^{\frac{r}{a}-1} e^{\sqrt{\frac{2a}{c^2}}(x-b)v-\frac{v^2}{2}} dv,
				\label{eq:fundamental F GBM}
    \end{equation}
		
    \begin{equation}
        G(x; a,b,c) = \int_0^\infty v^{\frac{r}{a}-1} e^{\sqrt{\frac{2a}{c^2}}(b-x)v-\frac{v^2}{2}} dv. 
				\label{eq:fundamental G GBM}
    \end{equation}

  Direct differentiation yields that $H'(x) > 0,$ $H''(x) > 0,$ $G'(x) < 0$ and $G''(x) > 0.$ Hence, both $H(x)$ and $G(x)$ are strictly positive and convex, and
they are, respectively, strictly increasing and decreasing. Without ambiguity in this section, we denote  $H(\delta) \equiv H(\delta; \kappa, \nu, \zeta)$ and $G(\delta) \equiv G(\delta; \kappa, \nu, \zeta)$ in this section. Alternatively, the functions $F$ and $G$ can be expressed as

\begin{align}
H(x;a,b,c) &= \exp\left(\frac{a}{2c^2}(x-b)^2\right)D_{-r/a} \left(\sqrt{\frac{2a}{c^2}}(b-x)\right), \\
G(x;a,b,c) &= \exp\left(\frac{a}{2c^2}(x-b)^2\right)D_{-r/a} \left(\sqrt{\frac{2a}{c^2}}(x-b)\right).
\end{align}
Here,  the function $D_{\cdot}$ is also known as parabolic cylinder function or Weber function, whose properties are elaborated in detail by \cite{erdleylyi1953}. The functions $F$ and $G$ will play a crucial  role in the solutions for $V$ and $J.$ 

\begin{proposition}\label{thm:GBM certificate price}
     Under the Martingale Model   in \eqref{eq:S_t diffusion} and \eqref{eq:delta_t diffusion}, we have:
     \begin{enumerate}
\item After the shipping certificate is exercised, it is optimal to never liquidate   the grain, and the value function $J(S_t, \delta_t) =  S_t$, for $t\ge T$. 
\item The value of  the shipping certificate   is given by 
    \begin{equation}
		 V(S_t,\delta_t) = S_t + \frac{1}{\kappa+r}\left(\delta_t-\frac{H(\delta_t)}{H'(\delta^*)} -\widehat{\delta} + \frac{\kappa(\nu-\widehat{\delta})}{r}\right) \mathbf{1}\{\delta_t \geq \delta^* \} - c_1 \mathbf{1}\{\delta_t < \delta^*\},		\label{eq:GBM certificate value}
		\end{equation}
	for $t\ge T$,  where the unique optimal exercise  threshold $\delta^*$ solves  the equation
		\begin{equation}
		\delta^{*} = \frac{H(\delta^*)}{H'(\delta^*)}-c_1(\kappa+r) + \widehat{\delta}- \frac{\kappa(\nu-\widehat{\delta})}{r}.
                \label{eq:optimal deltastar}
		\end{equation}
\end{enumerate}
                 The optimal exercise and liquidation strategies are respectively given by
		\begin{align}
 		\tau^* = \inf \{\,t \geq T : \delta_t \leq \delta^{*} \,\}, \quad \text{ and } \quad 
		\eta^* = \infty.\\
 		\end{align}
\end{proposition}

\begin{figure}[t!]\begin{center}
    \includegraphics[width=3.8in,trim=0.9cm 0.3cm 1.7cm 0.8cm, clip=true]{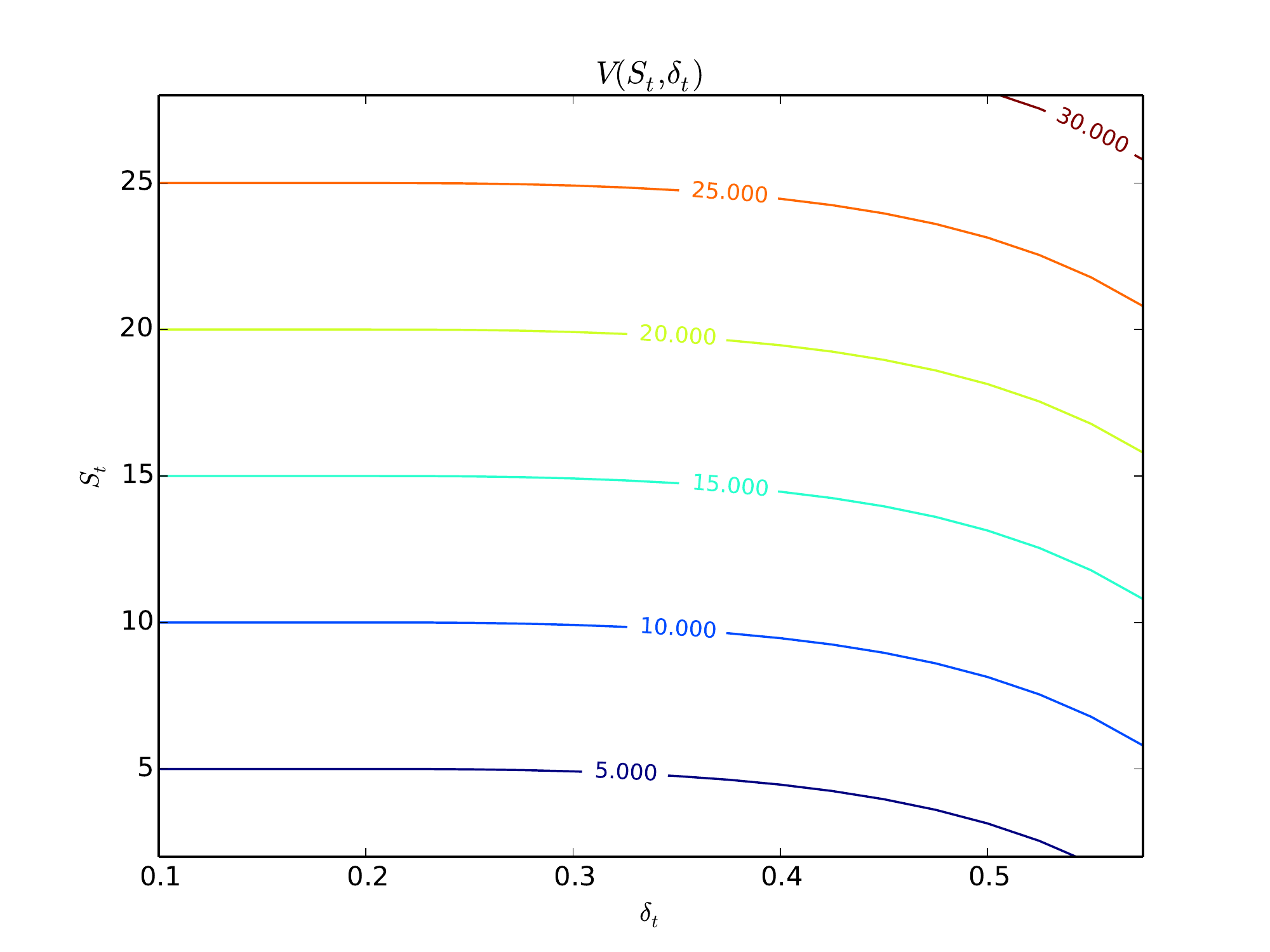}
    \caption{\small{The  shipping certificate price  $V(S_t,\delta_t)$  as a function of spot price $S_t$ and market  storage rate  $\delta_t$. Parameters are $r=0.03,$ $c_1 = 0,$ $c_2=0,$ $\kappa=0.3,$ $\zeta = 0.2,$ $\nu = 0.07,$ $\widehat{\delta}=0.06.$ The optimal exercise  level is $\delta^*=0.21.$}}
    \label{fig:GBM V prices}
    \end{center}
\end{figure}


We observe from \eqref{eq:GBM certificate value} that the shipping  certificate value  is separable in terms of the terminal spot price  $S_T$ and market storage rate  $\delta_T$ at time $T$. As we can see in  Figure \ref{fig:GBM V prices}, the  shipping certificate price $V(S_T, \delta_T)$ is increasing in both $S_T$ and $\delta_T$, and  always dominates $S_T.$  When the market  storage rate $\delta_T$ is below the critical level  $\delta^*$ at time $T$,  the shipping certificate price is equal to  the spot price $S_T$, implying   an immediate exercise by the holder. In this model, the non-convergence or basis is determined by the market storage rate $\delta_T$  since the exercise and liquidation strategies do not depend on $S_T$ at all. 

 In fact, the basis is roughly proportional to both the present value of the storage differential $\delta_T - \widehat{\delta}$ and the present value of the the average storage differential $\nu-\widehat{\delta}$. If $\delta_t<\widehat{\delta},$ then the value of the certificate decreases because it is currently cheaper to store at the real rate vs the certificate rate. Furthermore, if $\nu<\widehat{\delta},$ then the average storage rate is lower than the certificate rate, so the value of the shipping  certificate decreases. This model explicitly states that agent must consider both the long-run storage differential $\nu - \widehat{\delta}$ and the immediate storage differential $\delta_T - \widehat{\delta}$ in choosing his exercise strategy $\delta^*.$ In particular, when we set the parameters $\zeta = 0,$ $\kappa = 0$ and $\nu = \delta,$ the problem reduces to the case with a constant  market storage rate. In this case, the optimal exercise  level becomes $\delta^* = \widehat{\delta}-c_1 r,$ and the basis is completely linearly proportional to the storage differential $\delta - \widehat{\delta}$. After exercising the shipping certificate and thus receiving the grain,   there is no benefit to sell the grain early  ($\eta^*=\infty$). This  is due to the martingale property of the spot price.

According to Proposition \ref{thm:GBM certificate price}, the critical level $\delta^*$ determines  the time  $\tau^*$ to exercise the shipping certificate,  and thus,  plays a role in the  non-convergence of the  futures at maturity. Indeed, the  higher the critical level $\delta^*,$ the more likely the agent will exercise the shipping certificate at maturity,  resulting in zero non-convergence. In contrast, a low $\delta^*$ implies a high likelihood of non-convergence at maturity. In fact, when $c_1=0$, the basis $w(S_T,\delta_T) > 0$ if and only if $\delta_T > \delta^*$.

Furthermore,   the conditional probability that a $T$-futures contract expires with a strictly positive basis is given by  
\begin{equation}
    \mathbb{Q}(w(S_T,\delta_T)>0| \mathcal{F}_t) = 1-\Phi\left(z_{t,T}^*\right),
		\label{eq:prob of nonconvergence}
\end{equation}
where 
\begin{align}
\begin{split}
z_{t,T}^* &= \frac{\delta^{*} - \bar{\nu}_{t,T}}{\bar{\zeta}_{t,T}} \,,\\
\bar{\nu}_{t,T} &= \delta_t e^{-\kappa(T-t)} + \nu (1-e^{-\kappa(T-t)}) \,,\\
\bar{\zeta}_{t,T}^2 &= \frac{\zeta^2}{2\kappa} \left(1-e^{-2\kappa(T-t)}\right)\,,
\label{eq:barmu}
\end{split}
\end{align}
\noindent and   $\Phi$ is the  standard normal   cdf. In particular, the probability of a strictly positive basis depends solely on the current storage rate $\delta_t$ and the long run parameters of $(\delta_t)_{t\geq 0}.$ This probability is completely independent of the realization $S_t$ at any time $t$, or its driving parameters $r$ and $\sigma!$ Furthermore, since non-convergence at maturity is undesirable behavior, we would like to know precisely how the parameters of our model affect $\delta^*$. Differentiating \eqref{eq:optimal deltastar} with respect to $\widehat{\delta}$, $c_1$, and $\nu$, respectively, we obtain the sensitivity in each of  these parameters. 
\begin{align}
\frac{d \delta^*}{d\widehat{\delta}}  = h(\delta^*) \left(1 + \frac{\kappa}{r} \right) \geq 0, \quad    \frac{d \delta^*}{d \nu} = -h(\delta^*)\frac{\kappa}{r} \leq 0, \quad 
\frac{d \delta^*}{d c_1} = - h(\delta^*)\left(\kappa + r\right)  \leq 0, \label{derivs}
\end{align}
where we have defined \[h(\delta^*):= \frac{H'(\delta^*)^2}{H(\delta^*) H''(\delta^*)} \geq 0. \] The function  $h$ is positive because  $H$ is positive, increasing, and convex. Therefore, we deduce the properties of the optimal exercise threshold $\delta^*$.\\

\begin{corr}\label{corr:GBM optimal stopping level}
Under the Martingale Model defined in \eqref{eq:S_t diffusion} and \eqref{eq:delta_t diffusion}, the optimal stopping threshold $\delta^*$ is increasing with respect to $\widehat{\delta},$ but   decreasing with respect to $\nu$ and $c_1.$\\


\end{corr}


Having solved the certificate pricing problem, we proceed to  examine how the storage optionality   propagates   to futures prices. At time $T$, the  agent will exercise  if  storage rate $\delta_T$ is lower than  the critical level $\delta^*$. Therefore,  a higher $\delta^*$ increases the chance an agent will exercise the shipping certificate immediately upon the futures expiration. A higher certificate storage rate $\widehat{\delta}$ increases $\delta^*$ and hence lowers the probability of non-convergence. On the other hand, a higher average storage rate $\nu$ increases the probability of non-convergence. This occurs due to the incentive to store in the cheaper market: if the certificate storage rate is higher (resp. lower) than the market rate, then  the agent will exercise  sooner (resp. lower). 

When calculating the derivatives in \eqref{derivs},  the magnitude of each derivative is roughly proportional to $\kappa$, the rate of mean reversion of the storage rate $\delta_t$.  Indeed,  as $\kappa$ increases, the long run effect of $\nu$ dominates, acting as an amplifier on $\delta^*.$ Therefore, under higher $\kappa,$ if the storage differential $\nu-\widehat{\delta}$ is positive, the basis increases more, whereas if $\nu-\widehat{\delta}$ is negative, the basis increases less. Intuitively, since the value $c_1$ increases the agent's cost to exercise, it is expected, as seen in \eqref{derivs}, that $\delta^*$ is decreasing in $c_1$. Lastly,  after exercising,   the agent's liquidation timing  $\eta^*$ is trivial, so the liquidation cost $c_2$ does not affect the exercise level $\delta^*$.

Next, we compute the futures price using the shipping certificate price given in   Proposition \ref{thm:GBM certificate price}. It follows from the property of the OU process  that $\delta_T | \delta_t$ is normally distributed with parameters $\bar{\nu}_{t,T}$ and $\bar{\zeta}_{t,T}$ which are given in \eqref{eq:barmu}. Following the definition in  \eqref{eq:futures}, the futures price is  given by
\begin{align*}
\begin{split}
    F(t, S_t, \delta_t;T) =& \quad \mathbb{E}\left[ S_T | \mathcal{F}_t \right] + \frac{1}{\kappa + r} \mathbb{E}\left[\left(\delta_T - \frac{H(\delta_T)}{H(\delta^*)} - \widehat{\delta} + \frac{\kappa(\nu-\widehat{\delta})}{r} \right) \mathbf{1}\{\delta_T \geq \delta^*\}| \mathcal{F}_t \right] \\
    & - c_1 \mathbb{Q}(\delta_T < \delta^* | \mathcal{F}_t)\,.
\end{split}
\end{align*}
By computing  the   conditional truncated expectations of $S_T$ and $\delta_T$, we obtain an explicit formula for the futures price.

\newpage

\begin{corr} \label{thm:GBM futures prices}
Under the Martingale Model defined in \eqref{eq:S_t diffusion} and \eqref{eq:delta_t diffusion}, the  grain futures price   is given by

\begin{align}
F(t, S_t, \delta_t;T) =& \quad e^{r(T-t)}\left[S_t + \frac{\nu}{r}\left(1-e^{-r(T-t)}\right) + \frac{\delta_t-\nu}{\kappa+r}\left(1-e^{-(\kappa+r)(T-t)}\right)\right] \\
				& \quad + \frac{1}{\kappa+r} \left[ \bar{\nu}_{t,T} + \frac{\phi\left(z_{t,T}^*\right)}{1-\Phi\left(z_{t,T}^*\right)}\bar{\zeta}_{t,T} - \int_{z_{t,T}^*}^\infty \frac{H(\bar{\nu}_{t,T} + \bar{\zeta}_{t,T} u)}{H'(\delta^*)} \phi(u) du  \right. \\
				& \quad + \left.  \left(\frac{\kappa(\nu-\widehat{\delta})}{r}-\widehat{\delta} \right)\left(1-\Phi\left(z_{t,T}^*\right)\right) \right] -c_1 \Phi\left(z_{t,T}^*\right), \qquad t\le T,
\label{eq:Martingale Futures Price}
\end{align}
where $z_{t,T}^*,$ $\bar{\nu}_{t,T}$ and $\bar{\zeta}_{t,T}$ are given in \eqref{eq:barmu}, and $\phi$ and $\Phi$ are  the  standard normal  pdf and cdf respectively.\\
\end{corr}

 We note that like the shipping certificate prices, futures prices can be separated into an expectation that depends on $S_t$ and another involving $\delta_t$. Despite the separation, since     $\delta_t$ appears in the diffusion for $S_t$, the two stochastic factors are not independent. The futures price encapsulates  a number of components: (i) the risk-neutral expectation of the   future spot price; (ii)  expected future basis $w(S_T, \delta_T)$ (see \eqref{basis_def}); and (iii) expected future exercise cost $c_2.$ Thus, by accounting for the expected future basis resulting from the storage differential $\delta_t-\widehat{\delta},$ the futures price for all $t \leq T$ in a market with shipping certificates carries a premium over the     price of a futures contract that delivers just the grain at time $T$. We therefore demonstrate that anticipated future storage differentials can impact current futures prices, including  the contracts that are  far from expiry.

With an understanding on the  theoretical behavior of grain futures prices under the Martingale Model, we now calibrate to empirical data, and discuss the results  and economic implications. We obtain futures prices from 2004-2011 for CBOT corn, wheat and soybeans contracts using Bloomberg terminal. We obtain spot prices from 2004-2011 for CBOT corn, wheat, and soybeans from an average of daily sale prices of several Illinois grain depots.\footnote{\url{http://www.farmdoc.illinois.edu/MARKETING/INDEX.ASP}} We also obtain the empirical certificate storage rate $\widehat{\delta}$, quoted in $\$/bushel,$ from the CBOT website.\footnote{\url{http://www.cmegroup.com/rulebook/CBOT/II/$n$/$m$.pdf where $(n,m) \in \{(10,10),(11,11),(14,14)\}.$}}\footnote{As a result of the certificate rate $\widehat{\delta}$ increase in 2009, the size of the empirical basis became  smaller afterward. However, a large strictly positive basis can still manifest in the future if the market storage rate $\delta_t$ is significantly higher than the new certificate rate  $\widehat{\delta}.$}  For the interest rate in our model, we use the 3-month LIBOR rate observed on the same date. There are several quoted prices for spot grain, differing only in the quality of the grain. This quality option allows the short to choose the grade he wishes to deliver, subject to some prior fixed conversion multiplier of the settlement price. In order to obtain a single series of spot prices, for every time $t$ we use the then cheapest-to-deliver price as the spot price for grain. 

Recall that the basis $w(S_T, \delta_T)$ is the premium of the certificate price $V(T, S_T)$ over the spot price $S_T. $ at time $T$.  On the day $t=0,$ we have the empirical futures prices $(F_k)_{k = 0}^N$ with maturities $(T_k)_{k = 0}^N,$ and a known spot price $F_0 = S_0;$  we seek a model-consistent futures curve $\textbf{F}_k(\nu, \kappa, \zeta, \delta_0)$ for $k = 0, 1, \ldots, N$ which best fits the empirical futures prices, given model parameters $(\nu, \kappa, \zeta)$ and $\delta_0$, with the model futures prices given in \eqref{eq:Martingale Futures Price}. Under this setup, the best fit calibrated futures curve $\textbf{F}_k^*$ for $k = 0, 1, \ldots, N$ minimizes the sum of squared errors (SSE) between the empirical futures curve and the model futures curve. Furthermore, the best-fit parameters are defined to be $(\nu^*, \kappa^*, \zeta^*, \delta_0^*)$ the model parameters which achieve the best fit futures curve. The other exogenous parameters $(r, S_t, \widehat{\delta}, c_1, c_2)$ are directly determined via contract specifications. Precisely, the calibrated parameters and the resulting futures curve are found from 

    \begin{align}
        \begin{split}
            (\nu^*, \kappa^*, \zeta^*, \delta_0^*) &= \argmin_{\nu, \kappa, \zeta, \delta_0} \sum_{k=0}^N (F_k-\textbf{F}_k(\nu, \kappa, \zeta, \delta_0))^2 \\
            \textbf{F}_k^* &= \vec{\textbf{F}}(\nu^*, \kappa^*, \zeta^*, \delta_0^*) \qquad k = 0, 1, \ldots N.
        \end{split}
		\label{def:SSE}
    \end{align}

\begin{figure}[t!]
    \centering
    \includegraphics[height=2.4in, width=3.22in, trim=0.9cm 0.1cm 1.2cm 0.5cm, clip=true]{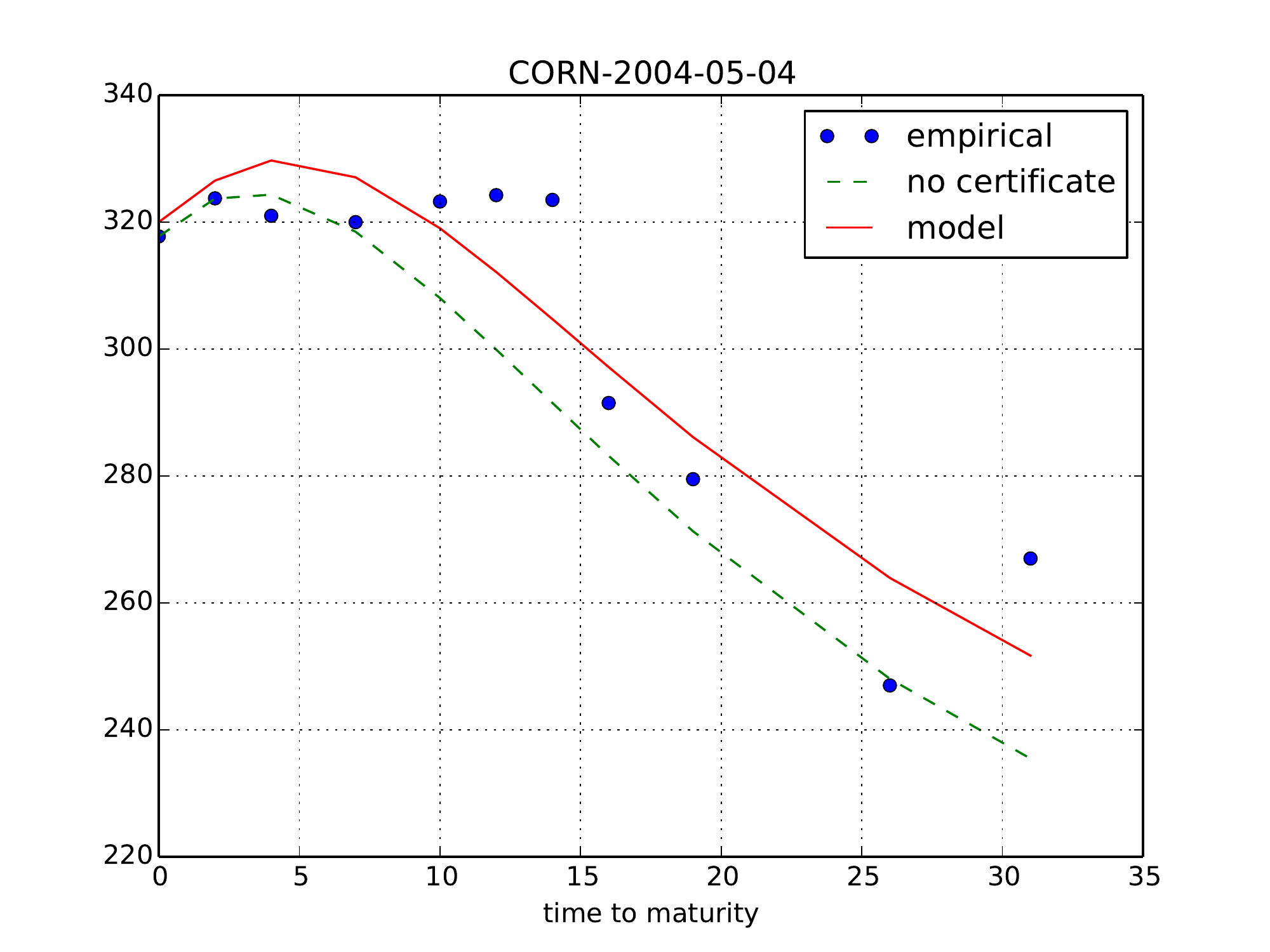}
    \includegraphics[height=2.4in, width=3.22in, trim=0.9cm 0.1cm 1.2cm 0.5cm, clip=true]{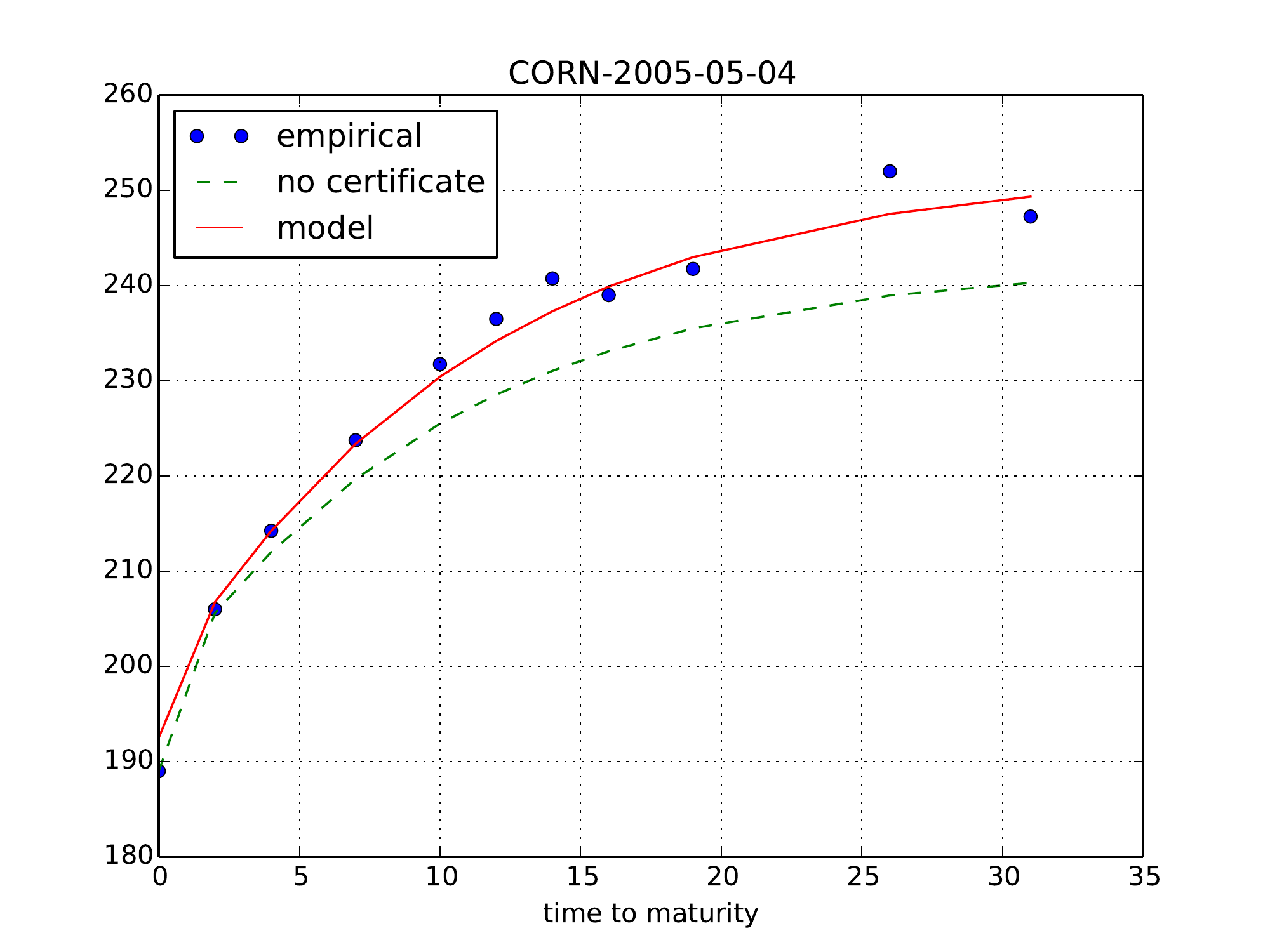}
    \caption{Calibrating the  Martingale Model to empirical  corn futures prices. The $x$-axis is time to maturity in months and the $y$-axis is the price of a bushel of corn in $cents.$ The `no certificate' curve is taken from equation \eqref{eq:GBM no timing options}. We use the fitted parameters from the `model' curve as inputs for the `no certificate' curve to illustrate the premium. Fitted parameters: (left) $\nu^* = -0.48,$ $\kappa^* = 0.0015,$ $\zeta^* = 0.0161,$ and $\delta_0^* = 1.2532$; (right)  $\nu^* = 0.832,$ $\kappa^* = 0.021,$ $\zeta^* = 0.428,$ and $\delta_0^* = 0.782.$ Other  parameters are $r=0.017,$ $S_0 = \{282, 167\}$ (cents), $\widehat{\delta} = 0.55$, and $c_1, c_2 = 0.$}
    \label{fig:GBM curve fit}
\end{figure}

In Figure \ref{fig:GBM curve fit}, we calibrate  the Martingale Model to the  corn futures prices on  two dates selected to show two characteristically different futures curves. On the left panel, the  futures curve   is downward sloping. With the expiring futures price and spot price being \$3.17 and \$2.81, respectively, a positive basis is observed. Intuitively, given that the current market  storage rate is higher than the certificate rate  ($\delta_0^* > \widehat{\delta}$),  the agent thus prefers the certificate storage rate over the market rate and will wait to exercise the certificate, resulting in  a positive basis.  The current storage rate $\delta_0^*$ is also  higher than the estimated  long-run  storage rate $\nu^*$. Therefore,   the model suggests that in the long run, a convenience yield will dominate, leading to a downward sloping futures curve. In contrast, the right panel also reflects a positive basis, but the futures curve is upward sloping. 

\begin{figure}[h!]
    \centering
    \includegraphics[height=2.4in, width=3.22in, trim=0.9cm 0.1cm 1.2cm 0.5cm, clip=true]{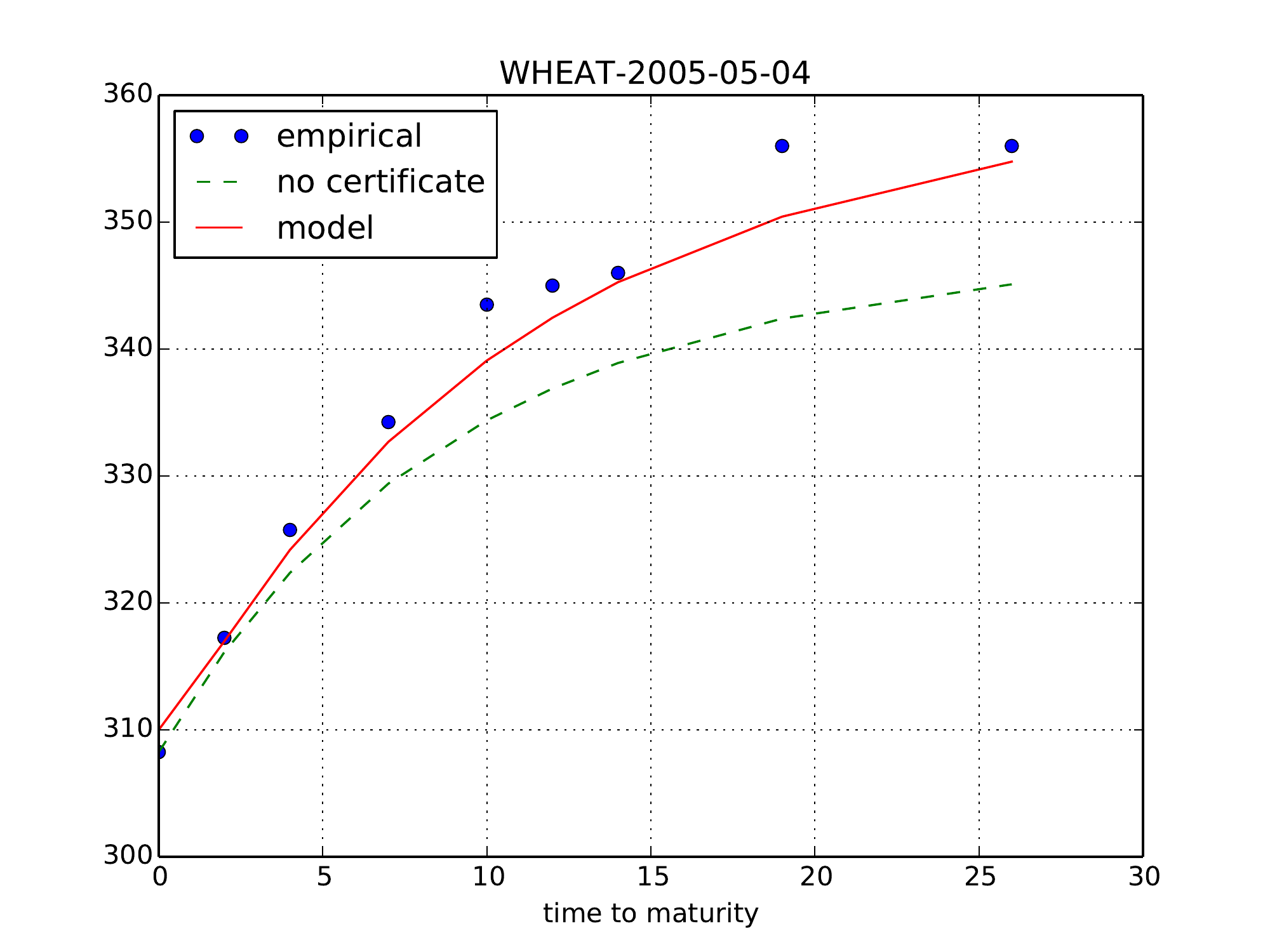}
    \includegraphics[height=2.4in, width=3.22in, trim=0.9cm 0.1cm 1.2cm 0.5cm, clip=true]{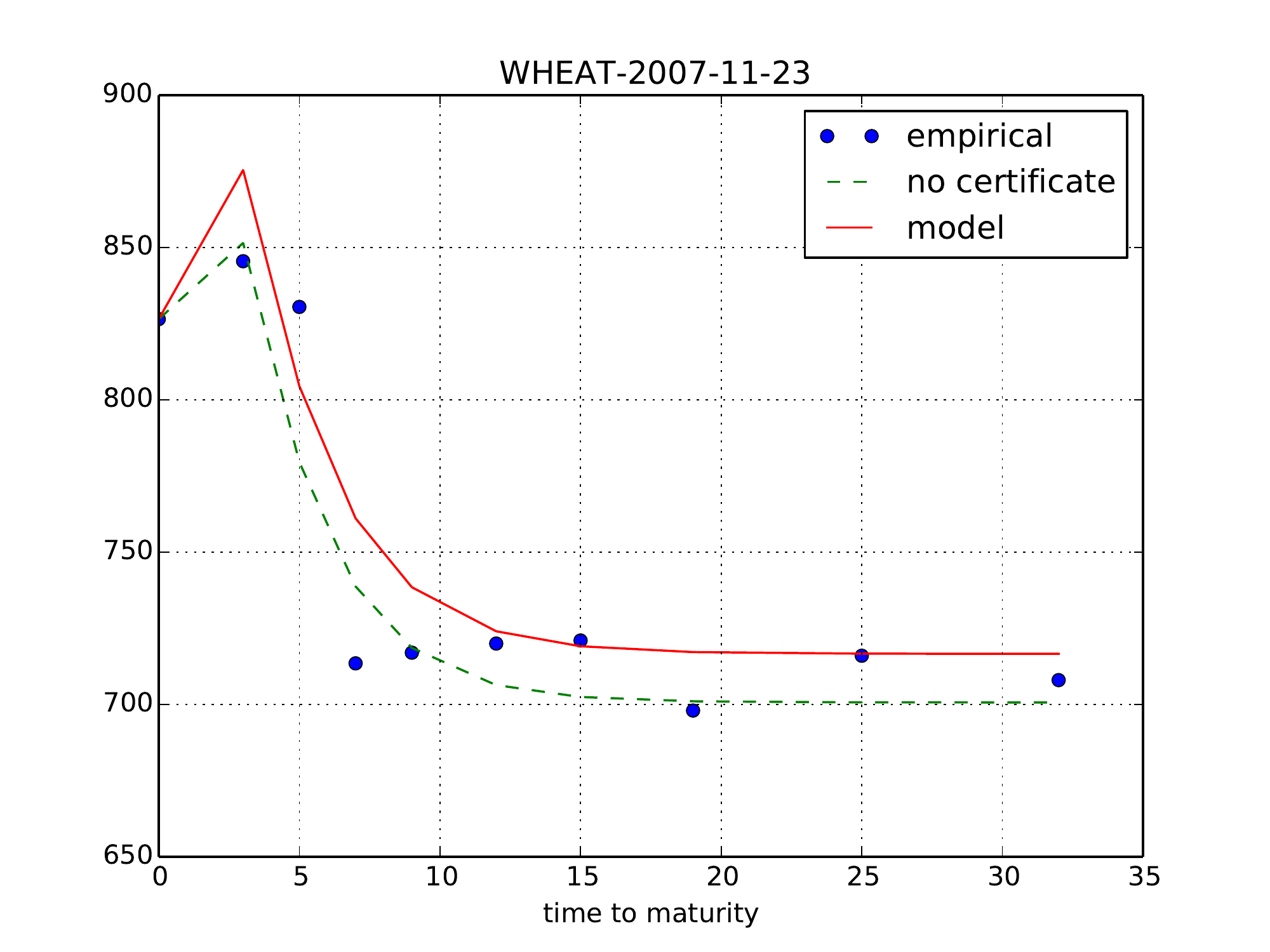}
    \caption{Calibrating the Martingale Model  to empirical wheat futures prices. The $x$-axis is time to maturity in months and the $y$-axis is the price of a bushel of wheat in $cents.$ The `no certificate' curve is taken from equation \eqref{eq:GBM no timing options}. Fitted parameters: (left) $\nu^* = 0.91,$ $\kappa^* = 0.035,$ $\zeta^* = 0.14,$ and $\delta_0^* = 0.871$; (right)  $\nu^* = -0.22,$ $\kappa^* = 0.0032,$ $\zeta^* = 0.61,$ and $\delta_0^* = 1.44.$ We use the fitted parameters from the `model' curve as inputs for the `no certificate' curve to illustrate the premium. Other parameters are $r=0.017,$ $S_0 = \{273 , 772\}$ (cents),  $\widehat{\delta} = 0.55$, and $c_1, c_2 = 0.$}
    \label{fig:GBM curve fit 2}
\end{figure}

 Figure \ref{fig:GBM curve fit 2} displays the calibrated futures curves under the Martingale Model for wheat on two dates when the futures market is in backwardation and contango, respectively. Again, non-convergence is observed on each of these two dates as the expiring futures price dominates the spot price.  On the left panel,  the futures curve is upward sloping  while the right panel shows that   the futures curve is downward sloping.  In this case,   the current market storage rate satisfies $\delta_0^*> \widehat{\delta}$, so it is optimal for the agent to continue to store at  the lower  certificate storage rate. Hence, the value of the associated timing option to exercise the shipping certificate yields  a positive basis. 

In addition, we consider the differences between the model futures curve and the futures curve generated without considering the timing options. To be precise, let the `no certificate' futures price $\psi(t, S_t, \delta_t; T)$ be
\begin{align}
\begin{split}
    \psi(t, S_t, \delta_t; T) &= \mathbb{E}[S_T|\mathcal{F}_t], \\
&= e^{r(T-t)}\left[S_t + \frac{\nu}{r}\left(1-e^{-r(T-t)}\right) + \frac{\delta_t-\nu}{\kappa+r}\left(1-e^{-(\kappa+r)(T-t)}\right)\right].
\label{eq:GBM no timing options}
\end{split}
\end{align} 
This follows from direct calculations and resembles the  first line of \eqref{eq:Martingale Futures Price}. In Figure \ref{fig:GBM curve fit}, we plot the values of $\psi(0,S_0,\delta_0^*; T_i)$ for $i = 0 \dots N,$ using the same fitted parameters from our model $(\nu^*, \kappa^*, \zeta^*, \delta_0^*)$ and the constraint that $S_t = F_0,$ the empirical terminal futures price. In other words, we ignore the data on spot grain prices, so there is initially zero basis, as would be the case under physical delivery, as opposed to receiving the  shipping certificate upon expiration.  As expected, the model futures prices with shipping certificate dominate the those without one, for all maturities.  As seen in  Figure \ref{fig:GBM curve fit},  the premium of the shipping certificate over the spot as the delivery item tends to be higher for longer maturities. Finally, the Martingale Model  fits both backwardated and upward-sloping futures curves well.

\clearpage
 
 Our model postulates  that the non-convergence occurs when the market storage rate is significantly higher than the certificate storage rate. Therefore, if  the two rates are brought into alignment, we expect that the basis to  diminish. Indeed, after years of high basis, the two exchanges,  CBOT and KCBOT,  enacted a series of reforms on the wheat futures to address the non-convergence phenomenon. During  February 2009  to  May 2011,  both exchanges instituted a one-time hike in the formerly constant storage rates, and subsequently adopted  a \emph{variable} certificate storage rate   for all wheat contracts, thus better aligning the certificate and market storage rates. For corn and soybeans, the exchanges merely raised the constant certificate storage rates once in Jan 2011. According to \eqref{eq:prob of nonconvergence},  these policy changes would decrease the likelihood of non-convergence. In Table \ref{tab:specs}, we compare  the average basis before and after the policy implementation  for each commodity. The average basis decreased by 6.57\% for  corn,   3.42\% for soybeans, and     2.29\% for wheat, suggesting that the effectiveness of changing  the certificate storage rate. In fact,  the study by \cite{Aulerich13}  finds  that introducing a {variable} certificate storage rate can significantly reduce non-convergence.\footnote{See also  the  CME Group  report, ``The Impact of Variable Storage Rates on Liquidity of the Deferred Month CBOT Wheat Futures" in 2010.}

\begin{table}[t!]\centering
\begin{small}
    \begin{tabular}{|l|cc|cc|}
\hline  Commodity & Average Basis (pre) & Max Basis (pre) & Average Basis (post) & Max Basis (post) \\
\hline\hline  Corn  & 9.11\%  & 27.52\% & 2.54\% & 13.50\% \\
        Soybeans  & 4.31\%  & 12.96\% & 0.89\% & 5.61\% \\
        Wheat     & 1.32\% & 14.10\% &  -0.97\% & 7.05\% \\
        \hline
    \end{tabular}\end{small}
    \caption{\small{Basis summary  during 2004-2014 before (pre) and after (post)   CBOT introduced   new  certificate policies  to facilitate convergence in the corn, soybeans, and wheat futures markets.  For corn and soybeans futures, the policy changed in  in January 2011. For wheat contracts, a variable storage rate policy was introduced in July 2010. For comparison across commodities, the basis here is computed in percentage at expiration according to  [(futures price/spot price)-1]$\times$100\%.}}
    \label{tab:specs}
\end{table}

\section{Local Stochastic Storage Model}\label{XOU with Local Storage}
We now consider an alternative to the Martingale Model. Since the commodity cannot necessarily be continuously traded, the market is incomplete, and one can specify a non-martingale evolution for  the   commodity price under the no-arbitrage risk-neutral measure. Commodities have the unique property that production can be increased or decreased in response to high or low prices, respectively. In addition, commodities can be consumed through production of end-goods (for example, turning corn into ethanol). In times of scarcity, production will increase to lower prices; in times of surplus, production will decrease while consumption continues to increase prices. Thus, the production and consumption process unique to commodities imply a mean-reverting price structure as suggested by \cite{Deaton96}.

Hence,  we propose an exponential OU (XOU) model for the  spot price. Under the risk-neutral measure  $\mathbb{Q}$,    the log-spot price of the  grain, denoted by  $U_t = \log S_t$, follows the OU process
\begin{equation}
d U_t = \alpha (\mu-U_t) dt + \sigma dW_t,
\label{model 1}
\end{equation}
where $W$ is standard Brownian motion under $\mathbb{Q},$ $\mu$ is the long-run mean, $\alpha$ is the rate of mean reversion, and  $\sigma$ is the volatility  of  the log-spot price. We impose the further regularity condition that $\sigma < \sqrt{2\alpha}.$

The market rate of storage $\delta_t$ is locally determined by the spot price through
\begin{equation}
    \delta_t = \beta U_t + \gamma.
\label{model 2}
\end{equation} 
We typically set the coefficient  $\beta \geq 0$ so that the storage cost increases linearly with the log-price of the commodity, with a possibly flat storage rate $\gamma>0$. In summary, we have described a local stochastic storage approach, whereby the market storage rate in \eqref{model 2} is a function of the stochastic spot price that follows the exponential OU model in \eqref{model 1}. Henceforth, we shall refer it as the XOU Model.


 Denote by $\mathbb{G} \equiv(\mathcal{G}_t)_{t \geq 0}$ the filtration generated by the log-spot price $(U_t)_{t\geq 0}$. Also, let  $\mathcal{S}$ be the set of all $\mathbb{G}$-stopping times, and $\mathcal{S}_{s,u}$ the set of $\mathbb{G}$-stopping times bounded by $[s,u]$. Note that  $\delta_t$ is a function of $U_t$, the optimal liquidation problem   is given by 
 \begin{equation}
    J(U_t) = \sup_{\eta \in \mathcal{S}_{t, \infty}} \mathbb{E}\left[e^{-r(\eta-t)}\left(\exp(U_\eta)-c_2\right) - \int_t^\eta \delta_u e^{-r(u-t)} du | \mathcal{G}_t \right],
	\label{eq:optimal liquidation}
\end{equation}
which applies  after the shipping certificate is exercised at time $\tau$. The   optimal timing problem to exercise the shipping certificate  is given  by
\begin{equation}
    V(U_T) = \sup_{\tau \in \mathcal{S}_{T,\infty}} \mathbb{E}\left[e^{-r(\tau-T)}\left(J(U_\tau)-c_1\right) - \int_T^\tau \widehat{\delta} e^{-r(u-T)} du | \mathcal{G}_T \right].
	\label{eq:optimal unloading XOU}
\end{equation}

  Since the shipping certificate serves as the delivery item instead of the actual grain,   the futures price $F(t,U_t ;T)$ at $t$ for the contract expiring at $T$ is given by 
\begin{equation}
    F(t,U_t;T) = \mathbb{E}[V(U_T) | \mathcal{G}_t].
\end{equation}

We now denote the operator from \eqref{eq:L}  by $\mathcal{L} \equiv \mathcal{L}^{\alpha, \mu, \sigma}$, which is the infinitesimal generator for the OU process $U$. To solve for the  certificate price and the agent's optimal policy,  we   solve the ODE
\begin{equation}
	    \mathcal{L}f(u) - rf(u) = 0,
\end{equation} 
which has the general solutions, $H(u; \alpha, \mu, \sigma)$ and $G(u; \alpha, \mu, \sigma),$ where $\mathcal{L},$ $H,$ and $G$ are defined in \eqref{eq:L}, \eqref{eq:fundamental F GBM} and \eqref{eq:fundamental G GBM}. In this section, without ambiguity, we denote  $H(u) \equiv H(u; \alpha, \mu, \sigma)$ and $G(u) \equiv G(u; \alpha, \mu, \sigma)$, both of which will play a role in the solution for $J$ and $V.$

\begin{proposition}\label{thm:XOU certificate prices}
    Under the XOU Model defined in \eqref{model 1} and \eqref{model 2}:
\begin{enumerate}
	\item The liquidation value is given by
		\begin{equation}
		J(u) = 
		\begin{cases}
		A H(u) - \frac{1}{\alpha+r}\left[\beta u + \gamma + \frac{\alpha(\beta\mu + \gamma)}{r} \right] &\text{ if $u<u^*$,} \\
		e^{u} - c_2 &\text{ if } u \geq u^*.
		\end{cases}
                \label{eq:XOU liquidation prices}
		\end{equation} 

	\item The certificate price is given by
		\begin{equation}
			V(u) = 
        \begin{cases}
        e^{u} -  c_1 - c_2 & \text{ if } u > \overline{u}^{**}, \\
        B H(u) + C G(u) - \frac{\widehat{\delta}}{r}              & \text{ if } \underline{u}^{**} \leq u \leq \overline{u}^{**}, \\
        A H(u) - \frac{1}{\alpha+r}\left[\beta u + \gamma + \frac{\alpha(\beta\mu + \gamma)}{r} \right]- c_1 & \text{ if } u < \underline{u}^{**},
        \end{cases}
        \label{eq:XOU certificate prices}
    \end{equation}
		
		\noindent where
		\begin{align}
		\begin{split}
		A &= \frac{e^{u^*} + \frac{\beta}{\alpha+r}}{H'(u^*)} \,,\\
		B &= \frac{e^{\overline{u}^{**}}G'(\underline{u}^{**})- \left(AH'(\underline{u}^{**})-\frac{\beta}{\alpha+r}\right)G'(\overline{u}^{**})}{H'(\overline{u}^{**})G'(\underline{u}^{**})-H'(\underline{u}^{**})G'(\overline{u}^{**})}\,, \\
		C &= \frac{e^{\overline{u}^{**}}H'(\underline{u}^{**})- \left(AH'(\underline{u}^{**})-\frac{\beta}{\alpha+r}\right)H'(\overline{u}^{**})}{H'(\underline{u}^{**})G'(\overline{u}^{**})-H'(\overline{u}^{**})G'(\underline{u}^{**})}\,,
		\end{split}
		\end{align}
		 and the optimal thresholds  $u^*$, $\overline{u}^{**}$ and $\underline{u}^{**}$ satisfy the equations:
		\begin{align}
		\begin{split}
		A H(u^*) - \frac{1}{\alpha+r}\left[\beta u^* + \gamma + \frac{\alpha(\beta\mu + \gamma)}{r} \right] &= e^{u^*} - c_2 \,,\\
                B H(\overline{u}^{**}) + C G(\overline{u}^{**})-\frac{\widehat{\delta}}{r} &= e^{\overline{u}^{**}}-c_1-c_2 \,,\\
                B H(\underline{u}^{**}) + C G(\underline{u}^{**})-\frac{\widehat{\delta}}{r} &= A  H(\underline{u}^{**}) - \frac{1}{\alpha+r}\left[\beta \underline{u}^{**} + \gamma + \frac{\alpha(\beta\mu + \gamma)}{r} \right]- c_1.		
		\end{split}
		\end{align}
 The optimal exercise and liquidation times,  respectively, are given by
		\begin{align}
		\begin{split}
		\tau^* &= \inf \{\,t\geq T : U_t \leq \underline{u}^{**}\, \text{ or }\, U_t \geq \overline{u}^{**} \,\} \,,\\
		\eta^* &= \inf \{\,t\geq \tau^* : U_t \geq u^* \,\}\,.\\
		\end{split}
		\end{align}
\end{enumerate}
\end{proposition}

The liquidation value $J(u)$ is entirely determined by the critical level $u^*$ at which the agent will sell the physical grain.  When the log-spot price $U_t$ surpasses $u^*$, both the storage cost $\delta_t$ and   spot price $S_t$ will be high. Since the asset price is mean-reverting, intuitively there is a potential  advantage to early liquidation before the asset reverts back to a lower value. In the holding region $\{u<u^*\}$ corresponding to low spot prices, the agent pays the present value of the storage rate $\beta u + \gamma,$ and the present value of the average storage rate $\beta \mu + \gamma.$ The conflict between increasing spot prices and higher storage rates, both of which are driven by $U_t,$ determines when the agent liquidates. Thus, both instantaneous storage rates and the long-run storage rates affect the certificate price. 

On the other hand, the certificate value $V(u)$ is determined by $two$ stopping levels: the optimal exercise threshold $\underline{u}^{**}$ and the optimal liquidation threshold $\overline{u}^{**}.$ When the spot price surpasses $\overline{u}^{**}$, the agent exercises $and$ liquidates to take advantage of temporarily higher spot prices, while avoiding higher storage rates. On the other hand, when the spot price decreases below $\underline{u}^{**},$ the agent exercises but does $not$ liquidate, because he wants to take advantage of a temporarily lower storage rate, storing in the real market at rate $\delta_t$ instead of at the certificate rate $\widehat{\delta}.$ Recall that due to our specification of the optimal stopping times $\eta^*$ and $\tau^*,$ the stopping levels   satisfy  $\overline{u}^{**} \geq u^* \geq \underline{u}^{**}.$ 

\begin{figure}[t!]\begin{center}
   \includegraphics[width=3.22in,trim=0.9cm 0.1cm 1.2cm 0.5cm, clip=true]{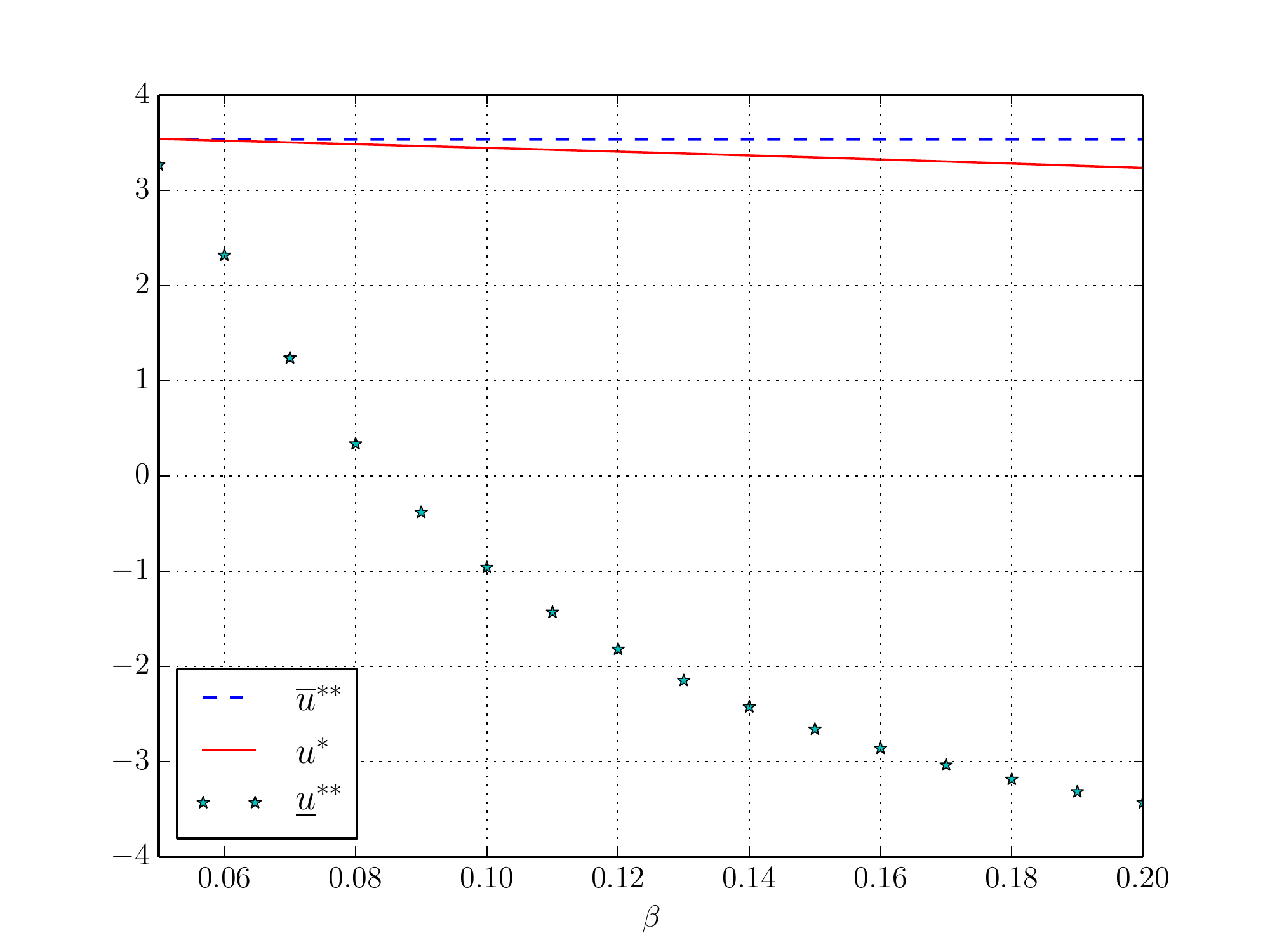}
   \includegraphics[width=3.22in,trim=0.9cm 0.1cm 1.2cm 0.5cm, clip=true]{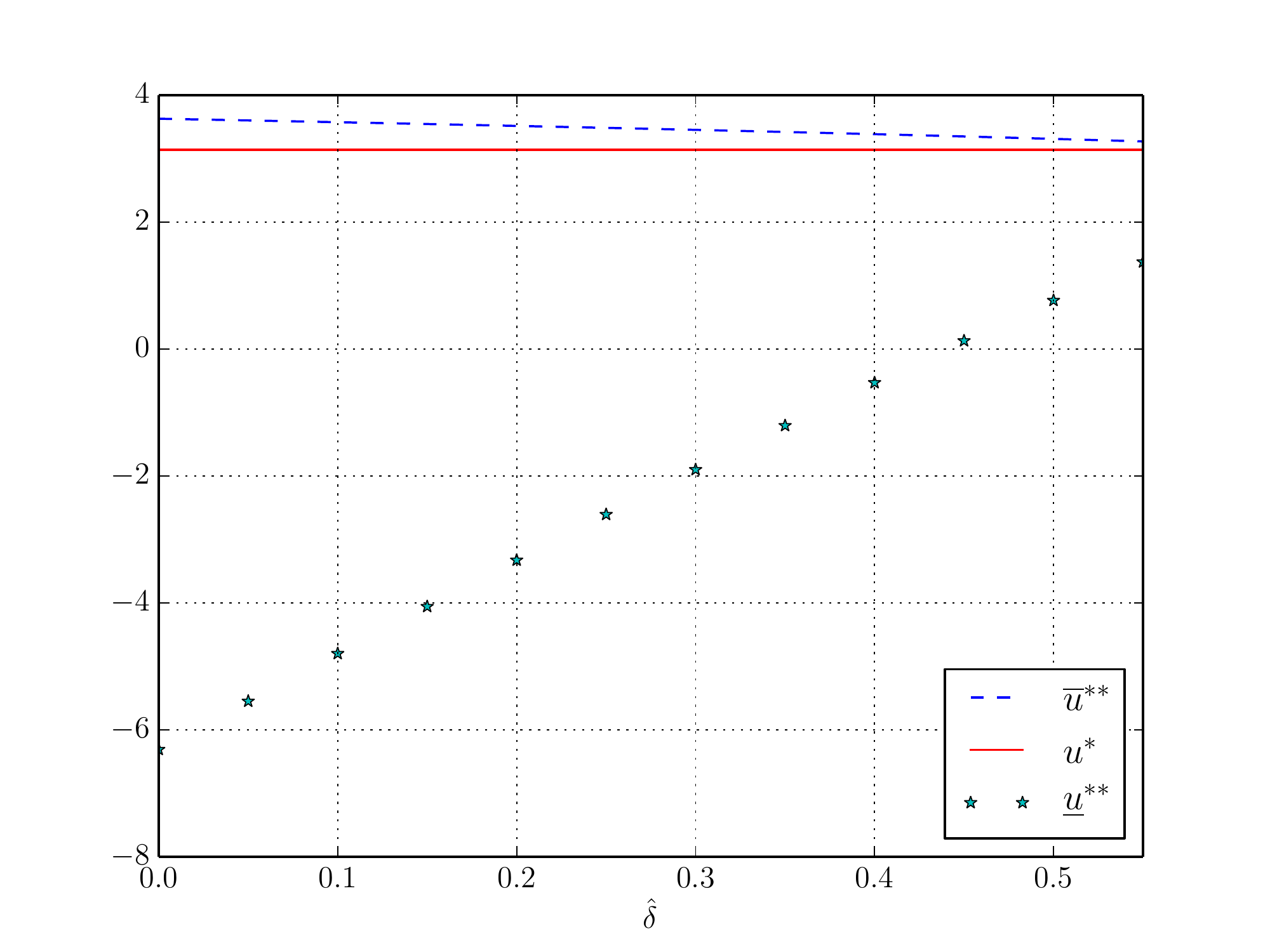}
   \caption{Optimal stopping levels for $[\underline{u}^{**}, u^*, \overline{u}^{**}]$, with default parameters $r=0.03,$ $U_t = \log 5,$ $\gamma =0, $ $\widehat{\delta} = 0.2,$ $\beta=0.08,$ $c_1 = 0,$ $c_2=0,$ $\mu=\log{30}$, $\alpha=0.1$, $\sigma=0.2$. We vary the parameters $\beta$ and $\hat{\delta}$ respectively in these plots.}
    \label{fig:local V}
    \end{center}
\end{figure}
\begin{figure}[h!]\begin{center}
   \includegraphics[width=4in,trim=0.9cm 0cm 1.2cm 0.5cm, clip=true]{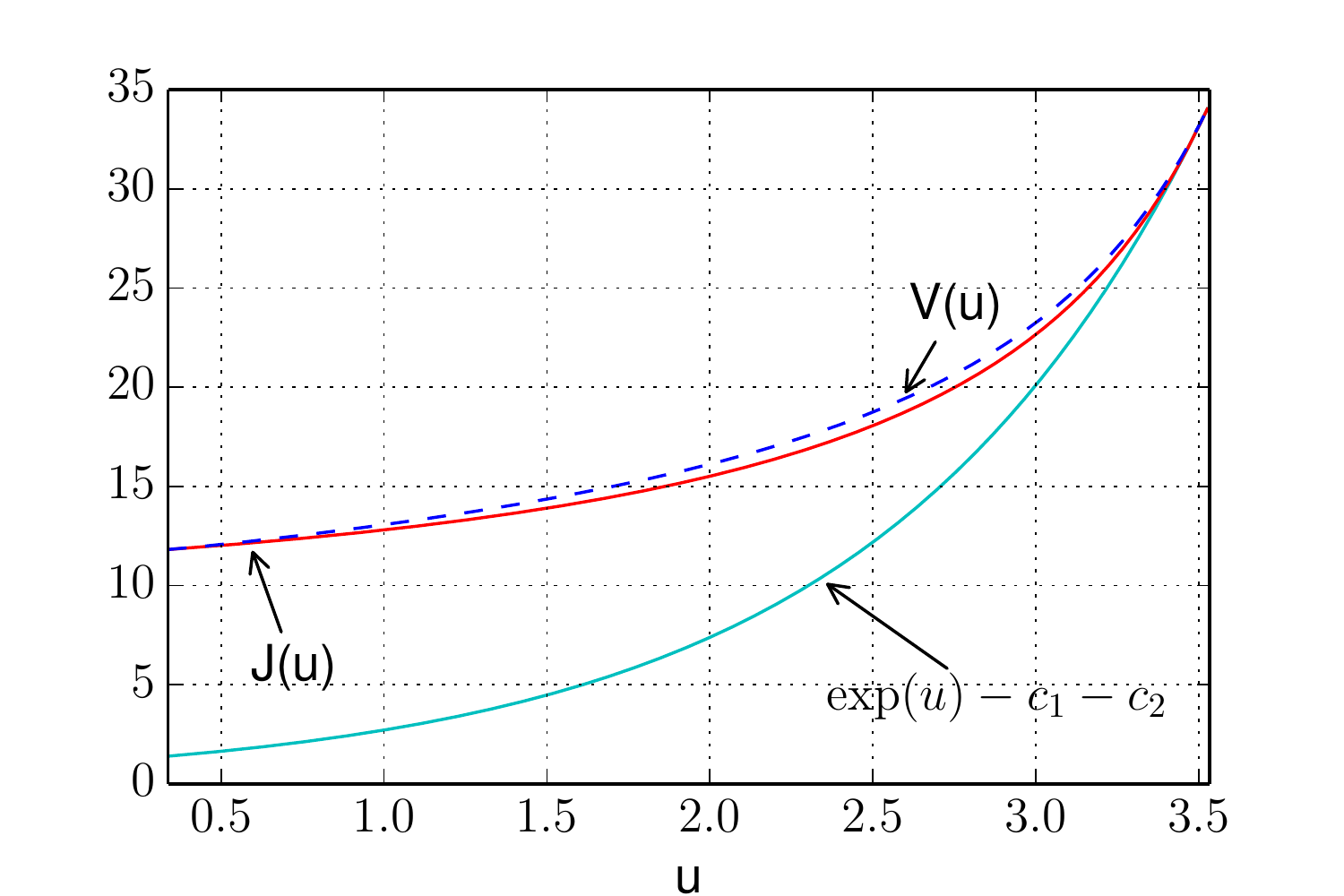}
    \caption{Immediate value $\exp(u)-c_1-c_2,$ liquidation value $J(u),$ and certificate price $V(u)$. The optimal stopping levels are given by $\underline{u}^{**}=0.337,$  $u^*= 3.485,$ $\overline{u}^{**} = 3.534.$ Parameters are $r=0.03,$ $\gamma =0, $ $\widehat{\delta} = 0.17,$ $\beta=0.10,$ $c_1 = 0,$ $c_2=0,$ $\mu=\log{30}$, $\alpha=0.1$, $\sigma=0.2$. }
    \label{fig:local V1}
    \end{center}
\end{figure}
\clearpage

Figure \ref{fig:local V} further illustrates the dependence of the three stopping levels $[\underline{u}^{**}, u^*, \overline{u}^{**}]$ on the parameters $\beta$ and $\widehat{\delta}.$ As $\beta$ increases, the gap between $\underline{u}^{**}$ and $\overline{u}^{**}$ increases, so   does  the basis. However, as $\beta \to 0,$  the market  storage rate goes to 0 in this example, inducing   the agent to exercise as soon as the futures expires and  store in the real market. Consequently, the region $\{\underline{u}^{**}, \overline{u}^{**}\}$ for holding the certificate vanishes, which also means that all three thresholds, $\underline{u}^{**}, u^*$, and $\overline{u}^{**}$ converge to the same value representing the optimal level to liquidate the grain.  A similar pattern is observed when the certificate storage rate $\widehat{\delta}$ increases since  the agent will again be incentivized to  exercise the shipping certificate immediately  to store in the real market.

Figure \ref{fig:local V1} also  reflects  the relationship among  the certificate price,   liquidation value, and     payoff from immediate exercise and liquidation. Note that \[V(u) \geq J(u) \geq \exp(u)-c_1-c_2,\] i.e. the shipping certificate price   dominates the liquidation value, which convexly dominates the immediate exercise value. The liquidation value is significantly higher than the immediate exercise value,  especially as $u \to -\infty.$ The value of $J(u)$ does not decrease as much as the immediate exercise value. Because the asset is mean reverting and the optionality is perpetual, the asset value is almost guaranteed to be eventually profitable. In this model, the ability to choose between two rates of storage adds merely a modest basis to the liquidation value. With the parameters in Figure \ref{fig:local V}, the maximum percent difference between $V(u)$ and $J(u)$ is 12.57\%. 

Recall that the basis $w(U_t)$ is defined as the difference between certificate and spot prices at maturity. As we established from a model-free argument, the basis $w(U_t) \geq 0.$ Therefore, a positive basis occurs when the agent chooses a strategy which is different than exercising and liquidating ($\eta^* > \tau > T$). In this scenario, the agent either waits to exercise because storage rates are too high to exercise and spot prices are too low to liquidate, or he has exercised to take advantage of lower storage prices but the spot price is too low to liquidate. In our model the probability of non-convergence depends completely on $\overline{u}^{**}.$ In particular, if $c_1=0$ and $c_2=0,$ then the basis $w(U_T) > 0$ iff $U_T < \overline{u}^{**}.$ Thus, the probability that there is a strictly positive basis at time $t$ for a contract maturing at $T \geq t$ is
\begin{equation}
    \mathbb{Q}\left(w(U_T)>0\,|\, \mathcal{G}_t\right) = \Phi\left(\frac{\overline{u}^{**} - \bar{\mu}_{t,T}}{\bar{\sigma}_{t,T}}\right),
\end{equation}
\noindent where 

\begin{align}
\begin{split}
\bar{\mu}_{t,T} &= U_t e^{-\alpha(T-t)} + \mu (1-e^{-\alpha(T-t)}), \\
\bar{\sigma}_{t,T}^2 &= \frac{\sigma^2}{2\alpha} (1-e^{-2\alpha(T-t)}).
\label{eq:barsigma}
\end{split}
\end{align}

Finally, in order to calibrate our model for empirical analysis, we derive futures prices by taking an expectation of the certificate prices.

Under the  OU model, the conditional log spot price $U_T | U_t$ is normally distributed with parameters $\bar{\mu}_{t,T}$ and $\bar{\sigma}_{t,T}$ given in \eqref{eq:barsigma}. The result then follows from computing  the associated conditional truncated expectations:
\begin{align}
\begin{split}
    & F(t,U_t; T) =  \quad \mathbb{E}\left[\left(e^{U_T} -  c_1 - c_2\right) \mathbf{1}\{U_T > \overline{u}^{**}\}| U_t \right] \\
    &\qquad\qquad  + \mathbb{E}\left[ \left( BH(U_T) + CG(U_T) - \frac{\widehat{\delta}}{r} \right)\mathbf{1}\{\underline{u}^{**} \leq U_T \leq \overline{u}^{**}\} | U_t \right] \\
    &\qquad\qquad  + \mathbb{E}\left[ \left( A H(u) - \frac{1}{\alpha+r}\left[\beta U_T + \gamma + \frac{\alpha(\beta\mu + \gamma)}{r} \right]- c_1 \right) \mathbf{1}\{U_T < \underline{u}^{**}\} | U_t \right]
\end{split}
\end{align}

\begin{corr}\label{thm:XOU futures prices}
The futures price $F(t,U_t; T)$ under the XOU Model defined in \eqref{model 1} and \eqref{model 2} is given by

\begin{align}
    F(t,U_t; T) &= \quad \exp\left(\bar{\mu}_{t,T}+\frac{\bar{\sigma}_{t,T}^2}{2} \right) \frac{\Phi\left(\bar{\sigma}_{t,T}-\overline{z}_{t,T}^{**}\right)}{1-\Phi\left(\overline{z}_{t,T}^{**} \right)}-(c_1+c_2)\left(1 - \Phi\left(\overline{z}_{t,T}^{**}\right)\right) \\
    & \quad + \int_{\underline{z}_{t,T}^{**}}^{\overline{z}_{t,T}^{**}} (B H(\bar{\mu}_{t,T} + \bar{\sigma}_{t,T}v) + C G(\bar{\mu}_{t,T} + \bar{\sigma}_{t,T}v))\phi(v) dv \\
    & \quad - \frac{\widehat{\delta}}{r}\left(\Phi\left(\overline{z}_{t,T}^{**}\right)-\Phi\left(\underline{z}_{t,T}^{**}\right)\right) + \int_{-\infty}^{\underline{z}_{t,T}^{**}} A H(\bar{\mu}_{t,T} + \bar{\sigma}_{t,T}v) dv  \\
    & \quad - \frac{\beta}{\alpha+r}\left(\bar{\mu}_{t,T}-\frac{\phi\left(\underline{z}_{t,T}^{**}\right)}{\Phi\left(\underline{z}_{t,T}^{**}\right)}\bar{\sigma}_{t,T}\right)-\left(\frac{\gamma}{\alpha+r}+\frac{\alpha(\beta\mu+\gamma)}{r(\alpha+r)}+c_1\right)\Phi\left(\underline{z}_{t,T}^{**}\right),
\end{align}

\noindent where 
\begin{align}
\overline{z}_{t,T}^{**}  = \frac{\overline{u}^{**} - \bar{\mu}_{t,T}}{\bar{\sigma}_{t,T}}, \qquad 
\underline{z}_{t,T}^{**} = \frac{\underline{u}^{**} - \bar{\mu}_{t,T}}{\bar{\sigma}_{t,T}},
\end{align}
  and $\bar{\mu}_{t,T},$ and $\bar{\sigma}_{t,T}$ are defined in \eqref{eq:barsigma}.
\end{corr}

 We calibrate our model futures curve to empirical corn, wheat and soybeans data and consider the accuracy and economic implications. Refer to Table \ref{tab:specs} for details of each contract's specification. Consider the futures curve at time $t=0.$ Recall that the best-fit futures curve can be defined in the following manner. Let the futures prices at time $T_k$ be $F_k,$ for maturity times $(T_k)_{k=0}^N,$ with $F_{0}= \exp(U_0) = S_0,$ so the futures price at $T_0=0$ is just the market quoted settlement price. Denote  the model futures curve generated at time $T_k$ by the parameters $(\beta, \gamma, \mu, \alpha, \sigma)$ be denoted $\textbf{F}_k(\beta, \gamma, \mu, \alpha, \sigma).$ 
 
 The best fit futures curve $\textbf{F}_k^*$ for $k = 0, 1, \ldots, N$ minimizes the weighted sum of squared errors (SSE) between the empirical futures curve and the model futures curve at time $t.$ Furthermore, the best-fit parameter is defined to be $(\beta^*, \gamma^*, \mu^*, \alpha^*, \sigma^*)$ the model parameters which achieve the best fit futures curve. The other exogenous parameters $(r, U_t, \widehat{\delta}, c_1, c_2)$ are directly determined via contract specifications (see Table \ref{tab:specs}). 
    \begin{align}
        \begin{split}
            (\beta^*, \gamma^*, \mu^*, \alpha^*, \sigma^*) &= \argmin_{\beta, \gamma, \mu, \alpha, \sigma} \sum_{k=0}^N (F_k-\textbf{F}_k(\beta, \gamma, \mu, \alpha, \sigma))^2 \\
            \textbf{F}_k^* &= \textbf{F}_k(\beta^*, \gamma^*, \mu^*, \alpha^*, \sigma^*) \qquad k = 0, 1, \ldots, N.
        \end{split}
		\label{def:SSE 2}
    \end{align}
    
\begin{figure}[t]
    \centering
    \includegraphics[width=3.22in]{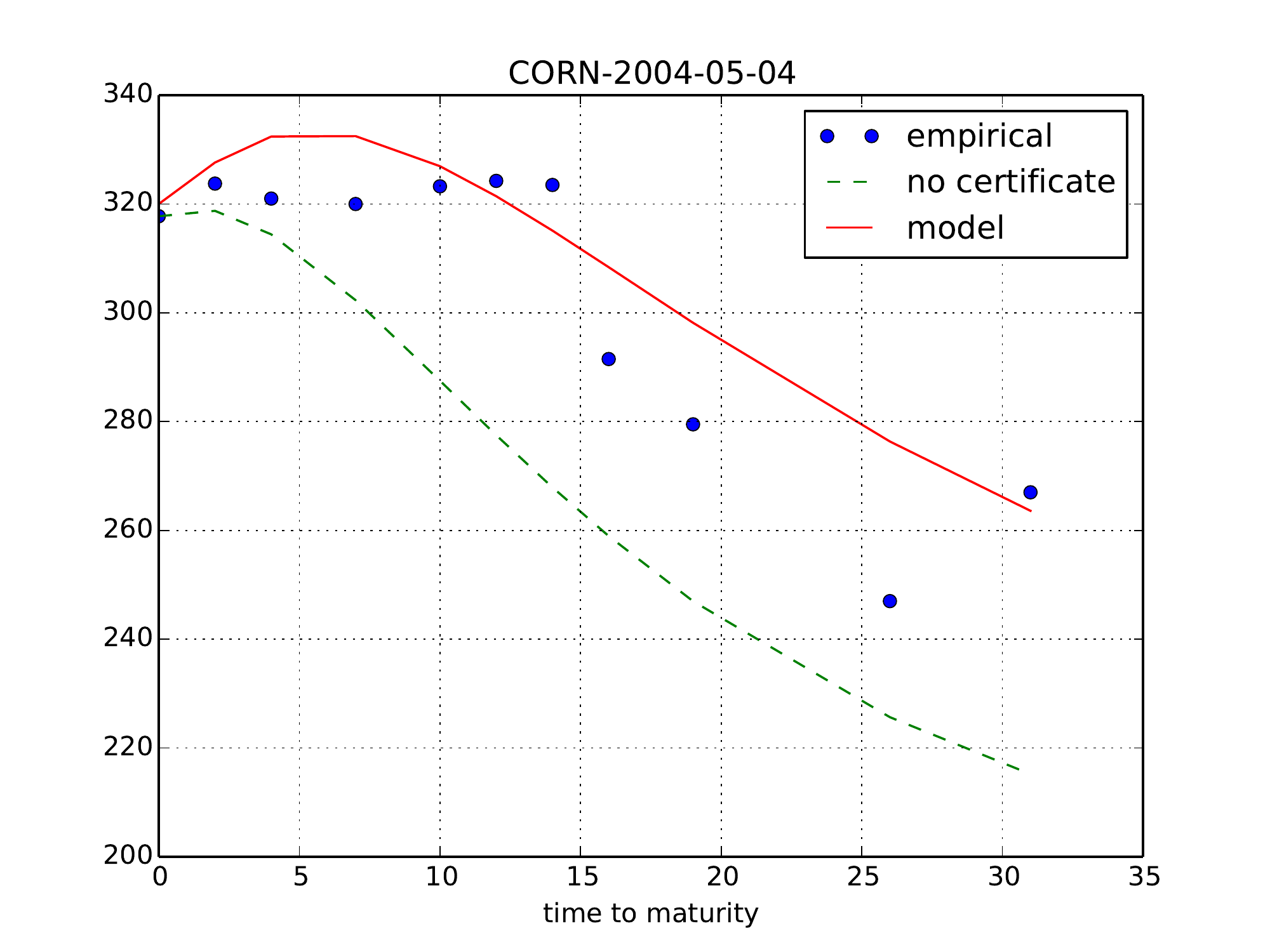}
    \includegraphics[width=3.22in]{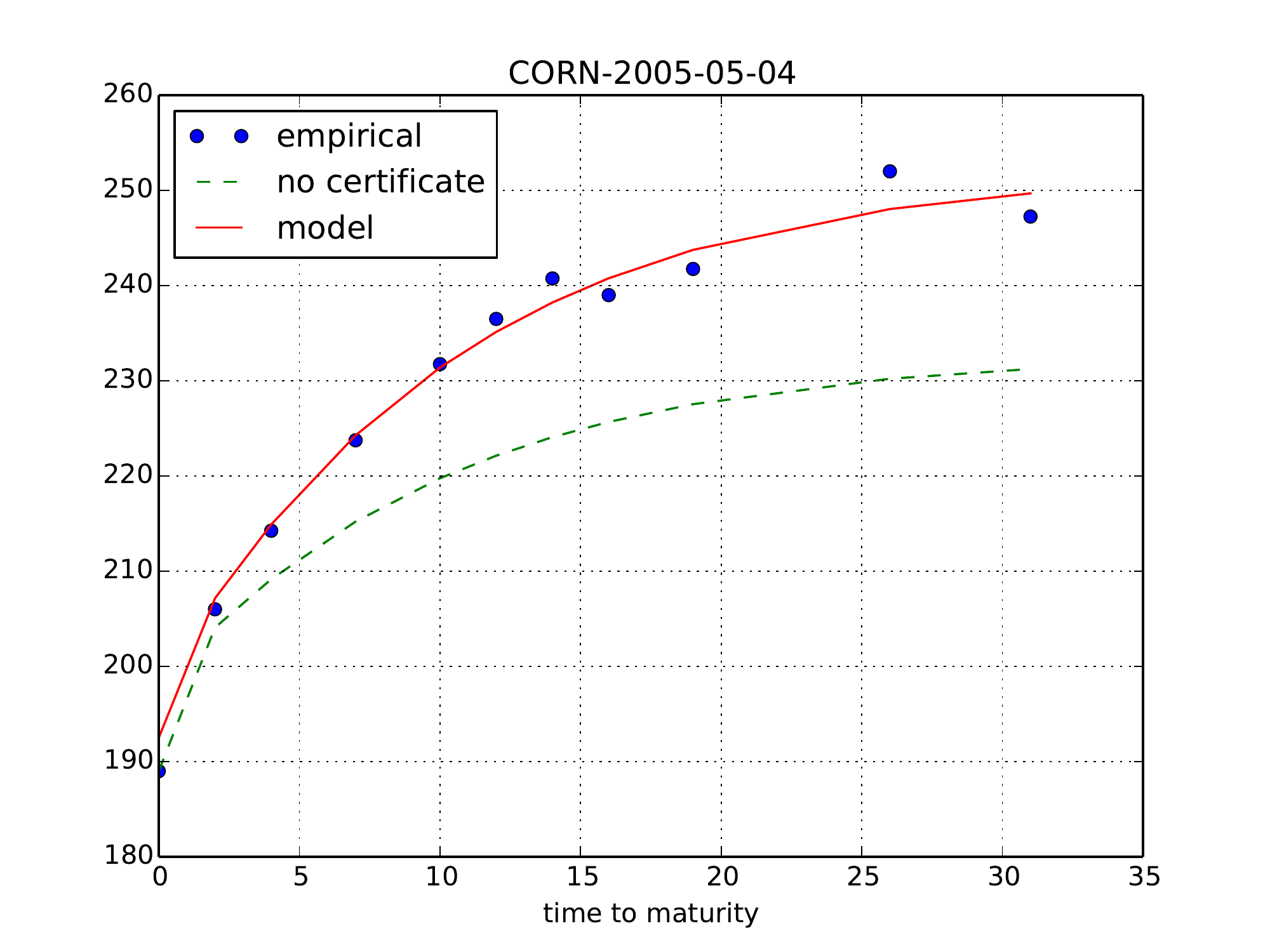}
    \caption{Calibrating the  XOU Model to  the empirical  corn futures prices. The $x$-axis is time to maturity in months and the $y$-axis is the price of a bushel of corn in $cents.$ The `no certificate' curve is taken from equation \eqref{eq:XOU local no timing options}. We use the fitted parameters from the `model' curve as inputs for the `no certificate' curve to illustrate the premium. Fitted parameters: (left) $\beta^* = 16.12,$ $\gamma^* = 15.08$ $\mu^* = 4.62,$ $\alpha^* = 0.058,$ and $\sigma^* = 0.40.$ Fitted parameters for the rightmost panel are $\beta^* = 10.75,$ $\gamma^* = 15.10,$ $\mu^* = 5.52,$ $\alpha^* = 0.10,$ and $\sigma^* = 0.12.$ Other  parameters are $r=0.017,$ and (in cents)  $S_0 = \exp(U_0) = \{281, 167\}$, $\delta_0^* = 106$, $\widehat{\delta} = 55$ and $c_1, c_2 = 0.$}
    \label{fig:XOU curve fit 2}
\end{figure}

The   $corn$ futures curves calibrated from  the XOU Model are illustrated in Figure \ref{fig:XOU curve fit 2}. We have selected the two dates to illustrate two characteristically different futures curves. On the left panel, the  futures curve that is upward sloping. With the expiring futures price and spot price being 317 and 281 (cents), respectively, a positive basis is observed. The current storage rate $\delta_0^* = 106.00,$ and the long run storage rate $\beta^* \mu^* + \gamma ^*= 89.55$ are both higher than the certificate rate $\widehat{\delta}.$  This storage rate spread   leads to  the   positive basis, while the long run storage rate anticipates a future   basis. Furthermore, the current spot price $U_0>\mu^*,$ which indicates  that the spot price will likely fall in the future, and results in   a downward-sloping futures curve. On the right panel, the futures curve is upward sloping and  the basis is more modest. The spot price current satisfies  $U_0 < \mu^*,$ so  the spot price is expected to rise in the future, generating a more contango futures curve.

\begin{figure}[h]
    \centering
    \includegraphics[height=2.4in, width=3.22in]{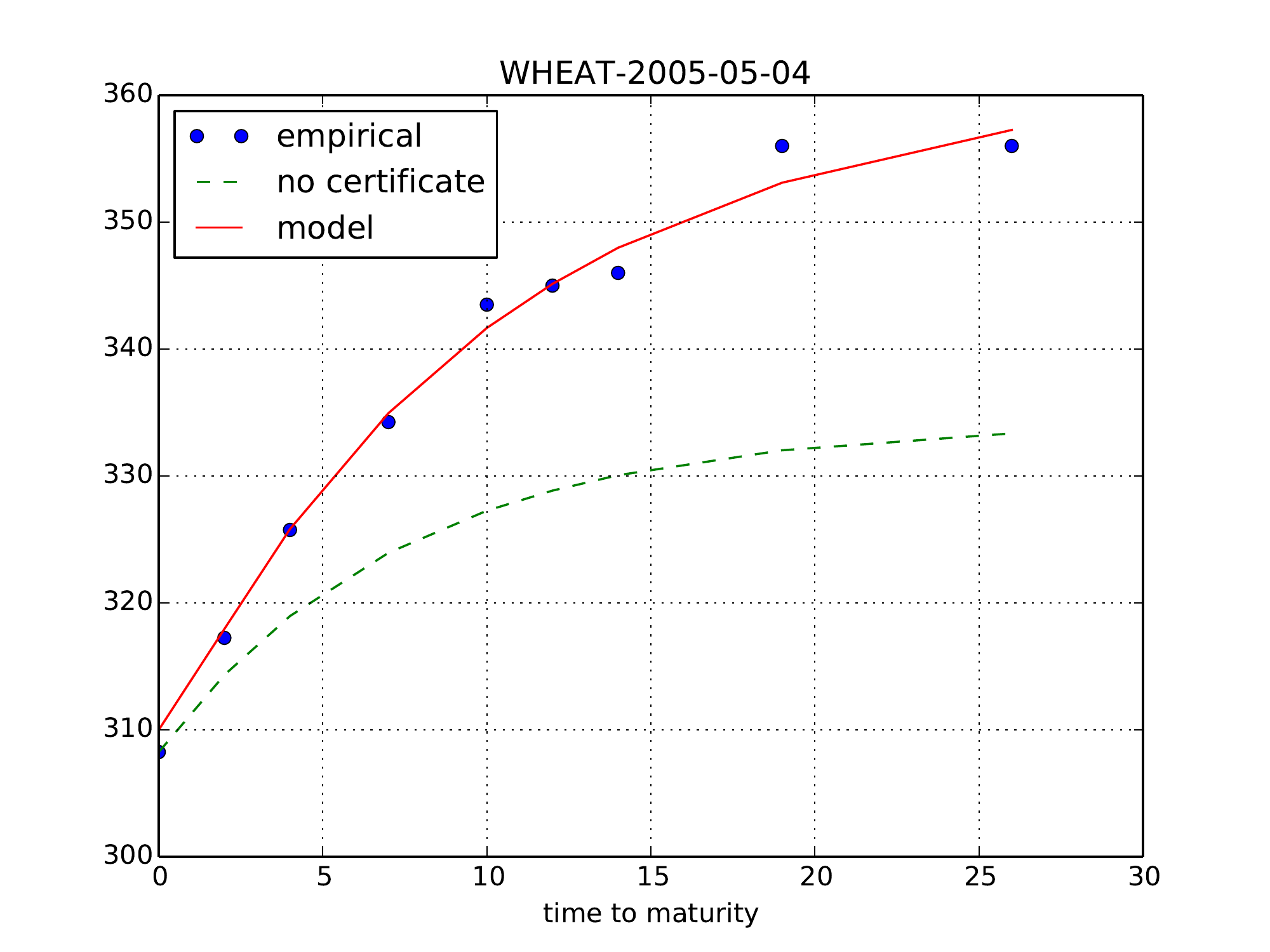}
    \includegraphics[height=2.4in, width=3.22in]{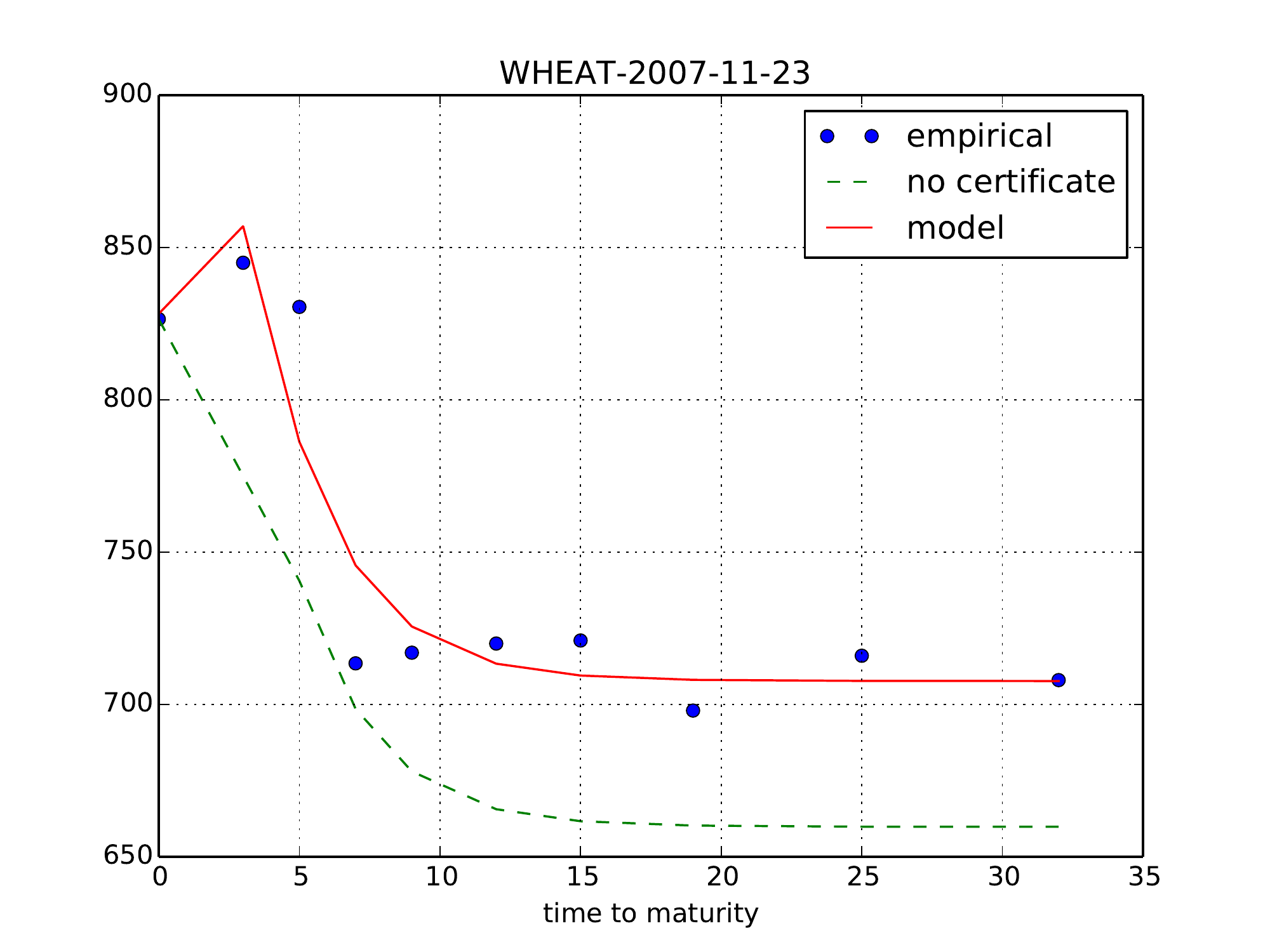}
    \caption{Calibrating XOU model with local storage to  the empirical  wheat futures curve. The $x$-axis is time to maturity in months and the $y$-axis is the price of a bushel of wheat in $cents.$ The `no certificate' curve is taken from equation \eqref{eq:XOU local no timing options}. We use the fitted parameters from the `model' curve as inputs for the `no certificate' curve to illustrate the premium. Fitted parameters: (left) $\beta^* = 9.98,$ $\gamma^* = 11.22$ $\mu^* = 5.83,$ $\alpha^* = 0.08,$ and $\sigma^* = 0.14.$ Fitted parameters for the rightmost panel are $\beta^* = 7.13,$ $\gamma = 14.16,$ $\mu^* = 5.90,$ $\alpha^* = 0.38,$ and $\sigma^* = 0.91.$ Other parameters are $r=0.017,$ $U_t = \{5.61, 6.65\},$ $\widehat{\delta} = 55$ and $c_1, c_2 = 0.$}
		    \label{fig:XOU local curve fit}
\end{figure}

In Figure \ref{fig:XOU local curve fit}, we see the results of our empirical calibration under the XOU model for  $wheat$  on two dates. The left panel shows a basis of around 12\%, and the futures curve is upward sloping. The current storage rate $\delta_0^* = 67.21,$ and the long run storage rate $\beta^* \mu^* + \gamma ^*= 69.84$ are both higher than the certificate rate $\widehat{\delta}.$  Therefore, the current storage rate leads to a positive basis, while the long run storage rate anticipates a future   basis. Furthermore, the current spot price $U_0<\mu^*,$ which indicates  that the spot price will likely rise in the future, and results in   an upward-sloping futures curve. In Figure \ref{fig:XOU local curve fit} (right), the basis is more modest at  5\%, while the futures curve is downward sloping. The current storage rate $\delta_0^* = 60.71,$ is higher than the certificate storage rate, but the long run storage rate $\beta^* \mu^* + \gamma^* = 53.87$ is lower than the certificate rate $\widehat{\delta}.$ Therefore, the current storage rate generates a smaller positive basis, while the long run storage rate anticipates little to  no basis on the futures curve. Furthermore, the current spot price is higher than the estimated long-run mean ($U_0 > \mu^*$),  and the  backwardated futures curve reflects the anticipation  that the spot price  will decrease in the future. This is consistent with the model's mean-reverting dynamics for the spot price.

In addition, we consider the differences between the model futures curve and the futures curve generated without considering the timing options. To be precise, let the `no certificate' futures curve $\psi(t, U_t, \delta_t; T)$ be

\begin{align}
\psi(t, U_t; T) &= \mathbb{E}[S_T|\mathcal{G}_t] \\
										&= \exp\bigg(e^{-\alpha(T-t)}U_t + \mu(1-e^{-\alpha(T-t)}) +    \frac{\sigma^2}{4\mu}(1- e^{-2\alpha(T-t)} )\bigg),
\label{eq:XOU local no timing options}
\end{align}
 which can be found in \citep[Sect. 2.2]{Li2016}. In Figure \ref{fig:XOU local curve fit}, we plot the futures curve, described by $\psi(0,U_0; T_i)$ for $i = 0 \dots N,$ using the same fitted parameters from our model $(\mu^*, \alpha^*, \sigma^*)$ and the initial assumption that $\exp(U_0) = F_0,$ the empirical terminal futures price. In other words, for the no-certificate case, we take the expiring futures price to be the spot price, and ignore the entire spot grain market prices. The last assumption means there is initially zero basis, as would be the case when physical grain is the delivery item.

\clearpage

As shown earlier, the model prices of futures of all maturities with a shipping certificate  dominate the respective contracts  without  one  due to the     timing options embedded in the shipping certificate. Furthermore, the difference increases as the futures maturity lengthens, indicating that the storage option exerts a more significant price impact  over a longer period of time.  In summary, we have shown that the timing options   in a shipping certificate  are a crucial component to explain the positive basis. As we have seen, the  exponential OU model is  able to capture forward anticipative behaviors of the basis and account for both backwardated and upwards-sloping futures curves. 

\begin{figure}[t!]
    \centering
    \includegraphics[height=2.4in, width=3.22in]{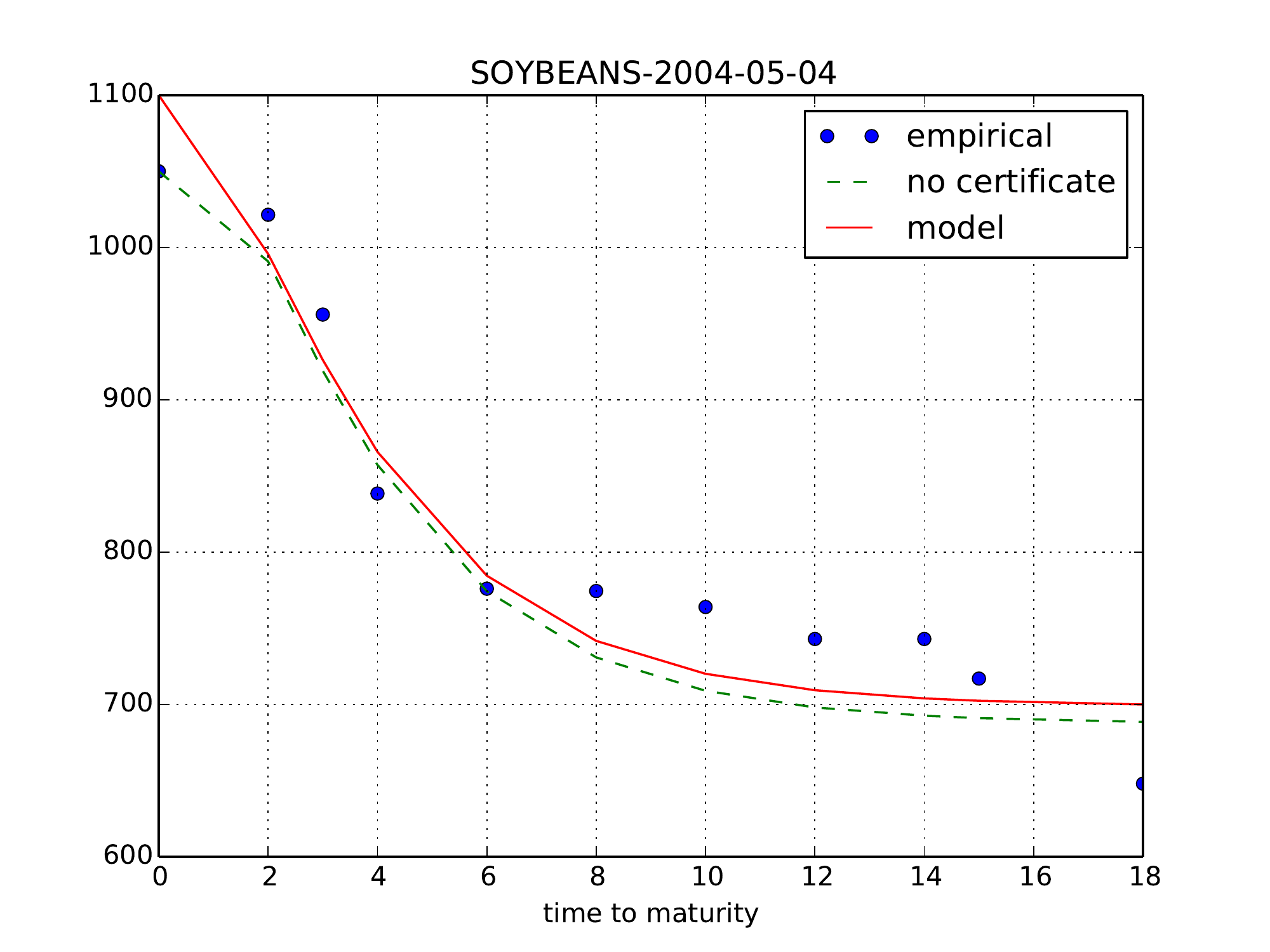}
    \includegraphics[height=2.4in, width=3.22in]{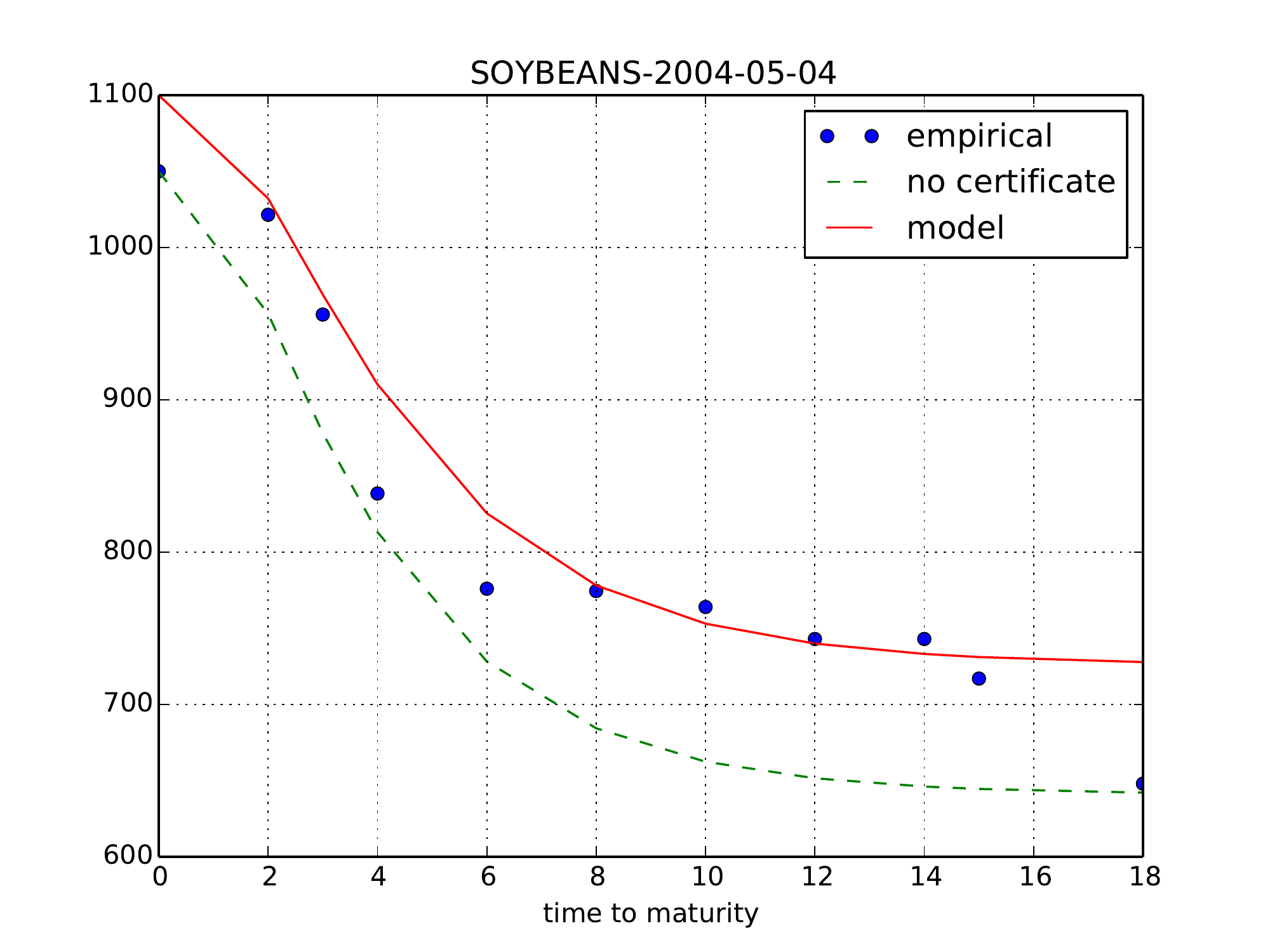}
    \caption{Calibrating  the Martingale Model (left) and  the XOU Model (right) to the empirical soybeans futures curve. The $x$-axis is time to maturity in months and the $y$-axis is the price of a bushel of wheat in $cents.$  The `no certificate' curve is taken from equations \eqref{eq:GBM no timing options} and \eqref{eq:XOU local no timing options} respectively. The fitted parameters from the `model' curve are used as inputs for the `no certificate' curve to illustrate the premium. Fitted parameters: (left) $\nu^* = -0.44,$ $\kappa^* = 0.023,$ $\zeta^* = 0.92,$ and $\delta_0^* = 1.42$; (right) $\beta^* = 11.88,$ $\gamma^* = 9.97$, $\mu^* = 5.91,$ $\alpha^* = 0.035,$ and $\sigma^* = 0.95.$ Other parameters are $r=0.017,$ $S_t = 1004.25,$ $\widehat{\delta} = 55$ and $c_1=c_2 = 0.$}
		    \label{fig:soybeans curve fit}
\end{figure}

We close this section by  comparing the empirical calibrations of the Martingale Model and the XOU Model in Figure \ref{fig:soybeans curve fit}. While both models are equally capable of estimating the immediate basis and fitting the empirical futures prices,   the value of the   timing option embedded in the shipping certificate is significantly higher  under the XOU Model than the Martingale Model. This can be seen from the spread between the  `model' curve (shipping certificate delivery) and the `no certificate' curve (physical spot delivery)  plotted on both panels. The `no certificate' curve generated from  the Martingale Model is   much closer to the fitted  `model' curve, whereas a visibly larger gap is observed in the XOU Model. Intuitively, the XOU Model tends to propagate the basis forward as the market storage rate is assumed to be positively correlated with the  spot price, but the Martingale Model assumes an independent stochastic (per bushel) storage rate. Therefore, the two models possess distinct features that address different market conditions, and have different implications to futures prices with longer maturities.


\section{Concluding Remarks}\label{Conclusion}

\begin{figure}[h]
    \centering
    \includegraphics[width=3.8in]{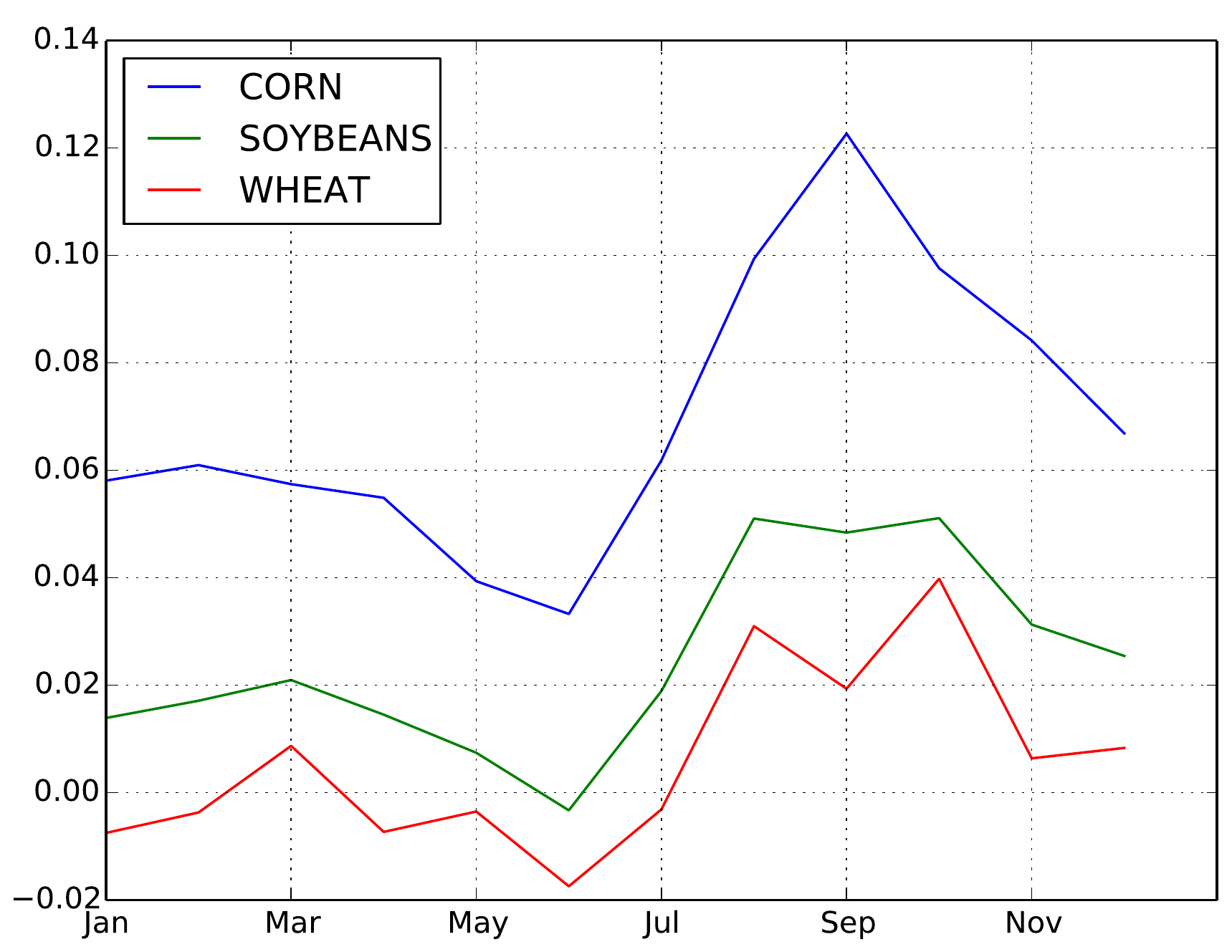}
    \caption{Seasonality of the average basis  for corn,     soybeans, and wheat during 2004--2010. The basis is highest during the harvest months (August-October) as storage rates are high due to grain silo capacity constraint. For instance, the average basis for soybeans  exceeds 12\% in September. From February to June,  the basis tends to be smaller since the  storage costs are lower due to empty grain silos before the next harvest begins.}
    \label{fig:seasonality}
\end{figure}

We have demonstrated   that the timing option embedded  in the shipping certificate for grains  leads to terminal non-convergence of futures and cash prices. The shipping certificate, by allowing its owner to choose the cheaper of two possible storage rates, therefore commands a premium over the physical grain itself. Our modeling approach captures the storage option of the shipping certificate by solving two optimal timing problems: one to determine the optimal liquidation strategy for physical grain and another for the optimal exercise strategy for the shipping certificate. 

We have proposed  two stochastic diffusion models for the spot grain and storage rate dynamics, one in which the storage rate process is OU and the spot price less storage costs is a martingale, and the other where  the spot price admits exponential OU dynamics along with a locally stochastic market  storage rate. Under both models,   explicit prices are provided  for the shipping certificate and associated futures curve. Furthermore, we  fit our models to empirical data during the periods of observed non-convergence. Our models not only capture the non-convergence phenomenon, but they also  demonstrate adequate  fit against the futures curve data when the market in backwardation or contango.

 In order to develop   tractable models with analytical solutions that are amenable to interpretation and calibration, we did not consider   the seasonality of grain prices, among other features. To  compare the basis over different months of the year, we illustrate  in  Figure \ref{fig:seasonality} the average basis for all three commodities. As shown, the  basis is typically higher during the fall harvest months (August to October) when available  storage capacity is low and market storage rates are high. This suggests that  the value of storage optionality, captured in our models here, is particularly high in these months.  In contrast, the basis is typically smaller from the winter through the summer while  the grain silos are being emptied before  the next harvest, and thus, storage rates are reduced  during the low season.

Overall,  both of our proposed  models are capable of generating model prices corresponding to  a variety of market situations, such as  high/low storage costs,  and backwardated/contango futures curves, and fit well for different commodities. Therefore, the proposed models seem to have sufficient components and strong economic rationale to reflect and quantify non-convergence in the grains markets. There is certainly room for incorporating additional characteristics, such as seasonality and other contractual features, such as quality and delivery options. However, the relatively small number of traded futures contracts  for each commodity may  limit the number of model parameters, and thus, model sophistication.

A better understanding of the price behaviors of commodity futures is also relevant to broader financial market, especially in the current era of so-called financialization of the commodity market (see   \cite{Tang12}), whereby commodity prices have become more correlated with the equity market. Moreover, commodity futures also play a role in the exchange-traded fund (ETF) market since most commodity ETFs are essentially dynamic portfolios of commodity futures; see \cite{GuoLeung,LeungWard}, among others. Therefore, for investors seeking spot exposure through commodity ETFs, any model which sheds light on the non-convergence phenomenon will affect investment decisions.  

While our models have  both  empirical explanatory power and theoretical foundation, our results do not rule out the possible scenario called the `failure of arbitrage'    in the grain markets, as suggested by the speculator hypothesis. Nevertheless,  alternative  theories of non-convergence can potentially be incorporated into our models.  The unique feature of our models is the embedded double timing option. This should motivate future research to investigate the valuation of such a timing option under different stochastic storage rate dynamics. Other   directions include adding to  futures   multiple   options, such as  the delivery option,   quality option, and    location option. Furthermore, we choose our models in this paper for analytical tractability which give closed-form certificate prices. One can also examine  certificate prices under more complex stochastic models, for example, with  stochastic interest rate, as well as stochastic volatility and jumps in the spot price or storage rate.

\section{Appendix}\label{Appendix}

\subsection{Proofs: GBM with Stochastic Storage}

\subsubsection{Proposition \ref{thm:GBM certificate price}}

We derive the certificate price by first determining the liquidation value function $J(S_\tau, \delta_\tau)$ and then substituting the value to solve the certificate problem $V(S_T, \delta_T).$ After applying the martingale  property of  $\left(M_t\right)_{t \geq 0} = \left(e^{-rt}S_t- \int_0^t \delta_u e^{-ru} du\right)_{t \geq 0}$, and applying the optional sampling theorem, the liquidation value function simplifies to
\begin{align}
\begin{split}
J(S_\tau, \delta_\tau) &= e^{r\tau } \left(\sup_{\eta \in \mathcal{T}_{\tau, \infty}} \mathbb{E}\left[e^{-r\eta}(S_{\eta}-c_2) - \int_\tau^\eta \delta_u e^{-ru} du | \mathcal{F}_{\tau} \right] \right) \\
								 &=e^{r\tau } \left(\sup_{\eta \in \mathcal{T}_{\tau, \infty}} \mathbb{E}\left[M_\eta + \int_0^\eta \delta_u e^{-ru} du - \int_\tau^\eta \delta_u e^{-ru} du - c_2 e^{-r\eta} | \mathcal{F}_{\tau}\right]\right) \\
								 &= e^{r\tau }\left(M_\tau + \int_0^\tau \delta_u e^{-ru} du + \sup_{\eta \in \mathcal{T}_{\tau, \infty}} \mathbb{E}\left[-c_2 e^{-r\eta} | \mathcal{F}_{\tau}\right] \right) \\
								&= S_\tau.
\end{split}
\end{align}
From the last step, we see that $\eta = \infty$ is optimal. Furthermore, after substituting $J(S_t, \delta_t) = S_t$ into the certificate pricing problem, and again using the fact that $M_t$ is a martingale, we obtain a solution for $V(S_T,\delta_T)$ which is separable in $S_T$ and $\delta_T:$

\begin{align}
 V(S_T,\delta_T) &= e^{rT} \left(\sup_{\tau \in \mathcal{T}_{T,\infty}} \mathbb{E}\left[e^{-r\tau} (S_\tau - c_1) - \int_T^\tau \widehat{\delta} e^{-ru} du | \mathcal{F}_{T} \right]\right)\notag \\
						&= e^{rT} \left(M_T + \int_0^T \delta_u e^{-ru} du + \sup_{\tau \in \mathcal{T}_{T,\infty}} \mathbb{E}\left[\int_T^\tau (\delta_u - \widehat{\delta}) e^{-ru} du -c_1 e^{-ru}| \mathcal{F}_T \right]\right) \notag\\
                                                &= S_T + \sup_{\tau \in \mathcal{T}_{T,\infty}} \mathbb{E}\left[\int_T^\tau (\delta_u- \widehat{\delta}) e^{-r(u-T)} du -c_1 e^{-r(\tau-T)} | \mathcal{F}_T \right]. \label{ww1}
 \end{align}
We denote the   second term  in \eqref{ww1} by  $P(\delta_T)$, where the function $P(\delta)$ satisfies the variational inequality

\begin{equation}
    \max\left\{\mathcal{L} P(\delta) - rP(\delta) + \delta - \widehat{\delta}, -P-c_1\right\} = 0,
\end{equation}

\noindent where $\mathcal{L} \equiv \mathcal{L}^{\kappa, \nu, \zeta}$ is the infinitesimal generator defined in \eqref{eq:L}. In order to determine $P(\delta)$ and the optimal stopping strategy $\tau^*,$ we first conjecture that $\tau^*$ takes the form \[\tau^* = \inf \{t \geq T : \delta_t \leq \delta^{*} \}\] for critical  stopping level $\delta^*$ to be determined. In other words, when the market storage rate $\delta_t$ is sufficiently small, the agent exercises to take advantage of the cheaper market storage rate, instead of  paying  the higher certificate rate $\widehat{\delta}.$ Thus, for $\delta_t > \delta^*$, we look for the solution of  the ODE $\mathcal{L} P(\delta) - rP(\delta) + \delta - \widehat{\delta}=0$, and for $\delta \leq \delta^*$ we require that $P(\delta) = -c_1$. This leads to the solution to the variational inequality
\begin{align}
P(\delta) = \left[A H(\delta) + \frac{1}{\kappa+r}\left(\delta-\widehat{\delta} + \frac{\kappa(\nu-\widehat{\delta})}{r}\right)\right] \mathbf{1}\{\delta  \geq \delta^*\} - c_1 \mathbf{1}\{\delta < \delta^*\},
\end{align}
along with the boundary conditions: $P(\delta^*) = -c_1$ and $P'(\delta^*) = 0$.  The latter is the  smooth pasting condition, which implies that  \[A = - \frac{1}{H'(\delta^*)(\kappa + r)}.\] 
  Enforcing these boundary conditions together also yields the optimal exercise level $\delta^*$  in  \eqref{eq:optimal deltastar}.  

 
Consider the function defined from \eqref{eq:optimal deltastar}
\begin{equation}
f(\delta^*) := \delta^* - \frac{H(\delta^*)}{H'(\delta^*)}.  
\end{equation}
First, the  properties of $H$ imply  that ${H}/{H'} \geq 0,$ so that $f(\delta^*) \leq \delta^*$. Taking the limit as $\delta^* \to -\infty,$ we have $f(-\infty) = -\infty.$ Furthermore, under the restriction $\zeta^2 \leq 2\kappa,$ and examining the terms inside the integrals of $H$ and $H'$, namely, 
\begin{align*}
H(\delta^*)  &= \int_0^\infty v^{\frac{r}{\kappa}-1} e^{\sqrt{\frac{2\kappa}{\zeta^2}}(x-\nu)v-\frac{v^2}{2}} dv, \\
H'(\delta^*) &= \int_0^\infty v^{\frac{r}{\kappa}} \sqrt{\frac{2\kappa}{\zeta^2}} e^{\sqrt{\frac{2\kappa}{\zeta^2}}(x-\nu)v-\frac{v^2}{2}} dv,
\end{align*}
we conclude  that $H/H' \leq 1.$ Therefore, $f(\delta^*) \geq \delta^*-1,$ and $f(\infty) = \infty.$ Finally, for $\delta^* \in (-\infty, \infty),$ $f' =  {H H'}/{H''} > 0$. Therefore, we have 
\begin{align}
    \lim_{\delta^*\to -\infty} f(\delta^*) = -\infty, \qquad \lim_{\delta^* \to \infty} f(\delta^*) = \infty, \qquad f'(\delta^*) > 0.
\end{align}
The  solution $\delta^*$ to \eqref{eq:optimal deltastar} is unique.

\subsection{Proofs: Exponential OU with Local Stochastic Storage}

\subsubsection{Proposition \ref{thm:XOU certificate prices}}

We consider a candidate interval type strategy for both $\tau$ and $\eta$. First, since the certificate price is monotonically increasing in the spot price, we consider  the optimal liquidation time $\eta^*$ to be of the form: $\eta^* = \inf \{t \geq \tau^* : U_t \geq u^*\}$.     In the liquidation problem represented by  $J$, we hold the commodity until the storage cost $\delta_t$, which is increasing in the commodity price $U_t,$ is sufficiently large relative to the commodity price. We solve a variational inequality for  the value function $J(u)$ and match the boundary condition at $u^*$ to get the solution. Assuming the conjectured form for $\eta^*,$ $J(u)$ satisfies  
\begin{equation}
        \begin{cases}
					\mathcal{L}J(u) - r J(u) = \beta u + \delta & \text{ if $u< u^{*}$}, \\
					J(u) = e^{u}-c_2              & \text{ if } u \geq u^*,
        \end{cases}
				\label{ode: XOU J proof}
\end{equation}

\noindent where $\mathcal{L} \equiv \mathcal{L}^{\alpha, \mu, \sigma}$ is the infinitesimal generator defined in \eqref{eq:L}. We apply the continuity and smooth pasting conditions to $J(u)$, and get 

\begin{align}
J(u^*) = e^{u^*}-c_2, \qquad J'(u^*) = e^{u^*},
\end{align}

This  gives the solution \eqref{eq:XOU liquidation prices} with \noindent $u^*$ satisfying \[f(u^*) = \frac{e^{u^*} + \beta/(\alpha+r)}{H'(u^*)} H(u^*) - \frac{1}{\alpha+r}\left[\beta u^* + \gamma + \frac{\alpha(\beta\mu + \gamma)}{r} \right] - e^{u^*} + c_2 = 0.\] When $ {\sigma}<\sqrt{2\alpha}$, the level $u^*$ admits a unique solution. First we can write \[f(u^*) = e^{u^*} \left(\frac{H(u^*)}{H'(u^*)}-1\right) + \frac{\beta}{\alpha+r}\left(\frac{H(u^*)}{H'(u^*)}-u^*\right) + C,\] for some constant C not depending on $u^*.$ Since $\frac{H}{H'} \leq 1$ then $f(u^*) \leq \frac{\beta}{\alpha+r}(1-u^*)+C$ so $\lim_{u^* \to \infty} f(u^*) = -\infty.$ Also, since $\frac{H}{H'} \geq 0,$ then $f(u^*) \geq -e^{u^*}-\frac{\beta}{\alpha+r}u^*+C$ so $\lim_{u^* \to -\infty} f(u^*) = \infty.$ Finally, we can look at \[f'(u^*) = -\frac{H(u^*)}{H'(u^*)}\left[e^{u^*}\left(\frac{H''(u^*)}{H'(u^*)}-1 \right) + \frac{\beta}{\alpha + r} \frac{H''(u^*)}{H'(u^*)} \right].\] Using similar arguments as from the previous appendix, we canshow $\frac{H''}{H'} \geq 1,$ so $f'(u^*) \leq -\frac{H(u^*)}{H'(u^*)} \frac{\alpha}{\alpha+r} < 0$ $\forall u^* \in \mathbb{R}.$ To recap, we have shown that

\begin{align}
    \lim_{u^*\to -\infty} f(u^*) = \infty, \qquad \lim_{u^* \to \infty} f(u^*) = -\infty, \qquad f'(u^*) < 0,
\end{align}

\noindent so our solution $u^*$ is unique.

On the other hand, in the exercise problem $V$,  the optimal strategy $\tau^*$ takes the form\[\tau^* = \inf \{t\geq T : U_t \leq \underline{u}^{**} \text{ or } U_t \geq \overline{u}^{**} \}, \] In other words, hold the commodity until either (i) the storage cost is   at or lower than $\underline{u}^{**}$ where the agent exercises, or (ii)  the commodity price is reaches the upper level $\overline{u}^{**}$ at which  the agent  exercises $and$ liquidates.   As such, the value function satisfies  
\begin{equation}
        \begin{cases}
					V(u) = e^{u}-c_1-c_2              & \text{ if $u>\overline{u}^{**}$}, \\
					\mathcal{L} V(u) - r V(u) = \widehat{\delta} & \text{ if $\underline{u}^{**} \leq u \leq \overline{u}^{**}$}, \\
					V(u) = A H(u) - \frac{1}{\alpha+r}\left[\beta u + \delta + \frac{\alpha(\beta\mu + \delta)}{r} \right]-c_1 & \text{ if $u<\underline{u}^{**}$}.
        \end{cases}
\end{equation}

\noindent The boundary conditions for $\underline{u}^{**}$ and $\overline{u}^{**}$ are 

\begin{align}
\begin{split}
V(\overline{u}^{**}) &= e^{\overline{u}^{**}}-c_1-c_2, \qquad V(\underline{u}^{**}) = J(\underline{u}^{**}) = A H(\underline{u}^{**}) - \frac{1}{\alpha+r}\left[\beta \underline{u}^{**} + \delta + \frac{\alpha(\beta\mu + \delta)}{r} \right]-c_1, \\
V'(\overline{u}^{**}) &= e^{\overline{u}^{**}}, \qquad \qquad \qquad V'(\underline{u}^{**}) = J'(\underline{u}^{**}) =  A H'(\underline{u}^{**}) - \frac{\beta}{\alpha+r}.
\end{split}
\end{align}
 We match the boundary conditions at $\underline{u}^{**}$ and $\overline{u}^{**}$ to get the solution \eqref{eq:XOU certificate prices}.

\linespread{0.1}

\begin{small} 
\bibliographystyle{apa}\linespread{0.1}

 \bibliography{mybib}
\end{small}
\end{document}